\documentclass[%
 reprint,
 longbibliography,
 amsmath,amssymb,
 aps,
pra,
]{revtex4-2}
\usepackage[english]{babel}
\usepackage{graphicx}
\usepackage{dcolumn}
\usepackage{bm}
\usepackage{xcolor}
\usepackage[colorlinks=True, linkcolor=blue,citecolor=blue,urlcolor=blue]{hyperref}
\usepackage[inline]{enumitem}
\usepackage{makecell}
\usepackage{placeins}

\begin{document}
\preprint{APS/123-QED}

\title{Robust gigahertz-range ac magnetometry with an ensemble of NV centers in diamond \\ using concatenated continuous dynamical decoupling}
\author{Takuya Kitamura$^1$}
\author{Genko Genov$^1$}
\author{Alon Salhov$^2$}
\author{Yutaka Kobayashi$^3$}
\author{Shinobu Onoda$^4$}
\author{Junichi Isoya$^5$}
\author{Alex Retzker$^{2,6}$}
\author{Fedor Jelezko$^1$}

 \affiliation{$^1$Institute for Quantum Optics, Ulm University, Albert-Einstein-Allee 11, 89081 Ulm, Germany}
 \affiliation{$^2$Racah Institute of Physics, The Hebrew University of Jerusalem, Jerusalem, 91904, Givat Ram, Israel}
 \affiliation{$^3$Advanced Materials Laboratory, Sumitomo Electric Industries Ltd., Itami, 664-0016, Japan}
\affiliation{$^4$Quantum Materials and Applications Research Center, National Institutes for Quantum Science and Technology, Takasaki, Gunma, 370-1292, Japan}
\affiliation{$^5$Faculty of Pure and Applied Sciences, University of Tsukuba, Tsukuba, Ibaraki, 305-8573, Japan}
\affiliation{$^6$AWS Center for Quantum Computing, Pasadena, California 91125, USA}

\begin{abstract}
Sub-picotesla level magnetometry has been demonstrated using negatively-charged nitrogen-vacancy (NV) centers in diamond by increasing the number of spins simultaneously used for sensing in an NV ensemble. However, such scale-up often introduces spatial inhomogeneities in detuning and control field amplitudes, which degrade sensitivity. Although several techniques have been utilized to overcome these challenges, including pulsed dynamical decoupling or shaped pulses, these are not generally compatible with the current state-of-the-art techniques for GHz-range AC magnetometry with NV ensembles, which are typically based on Rabi oscillations. In this work we experimentally demonstrate GHz-range AC magnetometry using a large ensemble of NV centers under spatially inhomogeneous drive fields by employing concatenated continuous dynamical decoupling, which is designed for robustness against such imperfections. We compare its performance with the conventional direct Rabi method and show that the robust dressed states in our method extend significantly the measuring range to weaker signals in GHz-range AC magnetometry. 
\end{abstract}

\maketitle

\section{
Introduction.
}
Quantum sensing \cite{degen_quantum_2017} with solid-state spins has been extensively studied following the development of single spin quantum systems \cite{gruber_scanning_1997}, and has broad potential applications in a wide range of scientific fields \cite{wolfowicz_quantum_2021, fang_quantum_2024}. Solid-state spins have been demonstrated to function as magnetic field sensors, most notably in color centers in diamond \cite{maze_nanoscale_2008, balasubramanian_nanoscale_2008}, silicon carbide (SiC) \cite{castelletto_quantum_2023}, and hexagonal boron nitride (hBN) \cite{healey_quantum_2023}. They have also been applied to measure magnetic fields with frequencies varying from DC to the GHz range. 
DC magnetometry
has been performed primarily via optically detected magnetic resonance (ODMR) \cite{balasubramanian_nanoscale_2008} or Ramsey-type pulsed measurements \cite{taylor_high-sensitivity_2008}, with applications in condensed matter physics \cite{casola_probing_2018} and biosensing \cite{le_sage_optical_2013}. 
AC magnetometry
in the kHz-MHz range has been accomplished by measurements based on Hahn echo \cite{maze_nanoscale_2008}, pulsed and continuous dynamical decoupling \cite{degen_quantum_2017}. It has also been applied in nano- and microscale nuclear magnetic resonance (NMR) \cite{allert_advances_2022}. 
While AC magnetometry with single spins in the GHz range has been demonstrated experimentally with a number of sensing protocols \cite{degen_quantum_2017,cai_long-lived_2012,stark_narrow-bandwidth_2017,ramsay_coherence_2023}, GHz sensing with spin ensembles has been challenging due to the inherent noise and inhomogeneities \cite{barry_sensitivity_2020}. Thus, it has been explored mainly by Rabi-type measurements \cite{wang_high-resolution_2015,appel_nanoscale_2015} and has been used in microwave device characterization \cite{horsley_microwave_2018, zhang_frequency_2024} and spin wave detection \cite{van_der_sar_nanometre-scale_2015, bertelli_magnetic_2020, carmiggelt_broadband_2023}. Importantly, solid-state quantum sensors have access to local near-field microwave magnetic fields at sub-mililimeter length scales, where further miniaturization of classical coils leads to self-resonant behavior and the coil no longer behaves as a simple inductive magnetic probe \cite{sidabras_extending_2019}.

Recently, highly sensitive magnetometry has been demonstrated using a large ensemble of NV centers, achieving sensitivities of several pT/$\sqrt{\mathrm{Hz}}$ \cite{wolf_subpicotesla_2015} and even reaching the sub-pT/$\sqrt{\mathrm{Hz}}$  \cite{barry_sensitive_2024},
but only for DC and AC sensing in the kHz range. Although using a higher number of spins enhances the signal strength, it becomes challenging to control all of them due to increased inhomogeneities, e.g., in the detuning or the amplitude of the control microwave (MW) field  \cite{barry_sensitivity_2020}.  
Several methods have been studied to overcome such inhomogeneities. Pulsed dynamical decoupling (PDD) is widely used for its robustness but is mainly suitable for sensing of AC magnetic fields in the kHz-MHz range \cite{naydenov_dynamical_2011, alvarez_performance_2010}. 
Pulse shaping techniques such as composite pulses \cite{levitt_composite_1986}, chirped pulses \cite{niemeyer_broadband_2013, genov_efficient_2020} and quantum optimal control \cite{scheuer_precise_2014, muller_one_2022, lim_efficiency_2025} have been intensively studied for better fidelity and can be integrated in pulsed measurements \cite{aiello_composite-pulse_2013, genov_efficient_2020}. However, the applicability of these methods to Rabi-type measurements for GHz-range magnetometry remains nontrivial, so 
sensitive magnetometry in the GHz range to date has primarily focused on the reduction of technical noise \cite{wang_picotesla_2022,alsid_solid-state_2023} and/or modulating the measured signal \cite{joas_quantum_2017}. 

Wang {\it et al.} achieved picotesla sensitivity by circumventing the DC noise of continuous-wave (CW) ODMR via modulation of the steady-state signal with an auxiliary microwave field \cite{wang_picotesla_2022}. Alsid {\it et al.} demonstrated picotesla sensitivity using the direct Rabi method by extensive optimization of the setup \cite{alsid_solid-state_2023}. However, limited research has been carried out on robust coherent control methods for GHz range AC magnetometry. While several PDD sequences (e.g., CPMG and XY8) have been applied to the GHz range \cite{joas_quantum_2017, meinel_heterodyne_2021}, they rely on phase flipping of the target signal, which is challenging in practical scenarios. Moreover, longitudinal RF dressed states have also been explored \cite{meinel_heterodyne_2021}, but their robustness under weak and inhomogeneous MW fields remains unclear.

In contrast, concatenated continuous dynamical decoupling (CCDD) \cite{cai_robust_2012, cohen_continuous_2017, farfurnik_experimental_2017, wang_coherence_2020, ramsay_coherence_2023,wang_observation_2021,teissier_hybrid_2017} is known to be robust against inhomogeneities and suitable for GHz-range magnetometry. 
GHz-range magnetometry with CCDD has been demonstrated with single NV centers in diamond \cite{stark_narrow-bandwidth_2017}, where amplitude noise across the temporal ensemble was effectively suppressed by introducing a second control field. This approach has been extended to enable precise phase measurement \cite{staudenmaier_phase-sensitive_2021}, and subsequent work has shown improved coherence times by optimizing the sequence based on correlations in the drive noise \cite{salhov_protecting_2024}. Nonetheless, much of the research up to now has been limited to single spins \cite{stark_narrow-bandwidth_2017, staudenmaier_phase-sensitive_2021, salhov_protecting_2024, salhov_2026_optimal}. For sensitive magnetometry, it is crucial to invesitgate the robustness of the CCDD sequence against amplitude noise across a spatial ensemble. Applications of CCDD to NV ensembles have been attempted under homogeneous control fields \cite{wang_coherence_2020}, but not yet in the context of GHz-range magnetometry. While CCDD has been applied to spin ensembles in hexagonal boron nitride (hBN), the spatial inhomogeneity of the driving field has been limited and the lifetime of the hBN electron spins is typically three orders of magnitude shorter than the one of NV centers \cite{patrickson_high_2024, patrickson_microwave_2025}. 

In this paper, we demonstrate GHz-range AC magnetometry with an ensemble of NV centers in diamond by employing concatenated continuous dynamical decoupling under inhomogeneous driving fields. The number of NV centers is increased by using high density of NV centers in diamond with a large laser excitation volume. Despite the inhomogeneity, we observe coherent oscillations driven by the target MW signal owing to the robustness of the CCDD. By comparison with the direct Rabi method, we show that CCDD magnetometry significantly improves the lower bound of measurable amplitude in GHz-range AC magnetometry. 

The paper is organized as follows. Section \ref{sec:scheme} outlines the principle of CCDD magnetometry. Section \ref{sec:CCDD} analyzes CCDD dynamics in the presence of inhomogeneous fields. Section \ref{sec:magnetometry} presents the experimental demonstration of GHz-range AC magnetometry with CCDD. Section \ref{sec:discussion} discusses the implications and limitations of the results, and Sec. \ref{sec:conclusion} concludes the paper. 

\section{Sensing scheme}\label{sec:scheme}
The dynamics of our ensemble spin system under CCDD can be described with the Hamiltonian ($\hbar=1$) \cite{salhov_protecting_2024,cai_robust_2012, stark_narrow-bandwidth_2017, genov_mixed_2019})
\begin{eqnarray}
H = 
&&\frac{1}{2} (\omega_0 + \delta) \sigma_z 
    + \Omega_1 (1 + \epsilon_1) \cos(\omega_0t) \sigma_x \nonumber \\
&&+ 2\Omega_2 (1 + \epsilon_2) \sin(\omega_0t) \cos (\widetilde{\Omega}_1t) \sigma_x \nonumber \\
&&+ \Omega_{\rm t}\cos(\omega_{\rm t} t + \xi) \sigma_x.
\label{eq1}
\end{eqnarray}
The first term represents the energy gap of the system $\omega_0$ and detuning $\delta$. 
The second and third term correspond to the two driving fields with Rabi frequencies $\Omega_k$, amplitude errors $\epsilon_k,k=1,2$ and modulation angular frequency $\widetilde{\Omega}_1$. The last term represents the MW signal that we try to measure, which we label the target signal. We note that $\delta$, $\epsilon_k$ can be time dependent.

\begin{figure}[t]
\includegraphics[width=8.6cm]{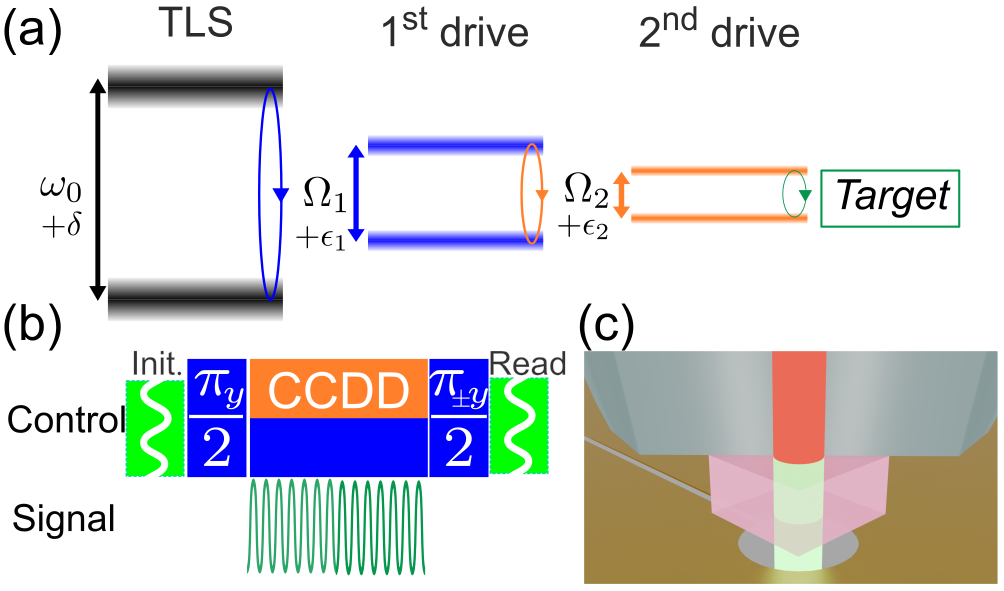}
\phantomsection
\caption{\label{fig_diagram} 
(a) Principle of GHz-range AC magnetometry with concatenated continuous dynamical decoupling (CCDD). In an ensemble of two level systems, the energy splitting $\omega_0$ is broadened by an inhomogeneous distribution of detuning $\delta$. A first continuous drive at Rabi frequency $\Omega_1$ is resonantly applied to suppress this detuning. In the rotating frame, the system acquires an effective splitting $\Omega_1$ in the dressed-state basis, where amplitude noise of the first drive $\Omega_1\epsilon_1$ appears as an effective detuning. To mitigate this noise, a second drive is concatenatedly applied at Rabi frequency $\Omega_2$. This doubly dressed state can be used to detect a target microwave signal when it is on resonance. 
(b) GHz-range AC magnetometry scheme using transverse CCDD. $\pi_y/2$ pulses are applied to prepare $x$ states, phase-locked to the first drive along $y$. 
(c) Schematic of the experimental setup around our ensemble system. An ensemble of NV centers inside a pink diamond is initialized by a green laser. Its spin state is controlled with a planar microwave (MW) antenna and readout through red-shifted photoluminescence (PL) collected by a compound parablocic concentrator (CPC).
}
\end{figure}

The limitation of direct Rabi magnetometry can be understood by considering 
the Hamiltonian of the system driven directly by a target signal in the rotating frame at $\omega_0$, and after applying the rotating wave approximation,
 
\begin{equation}\label{Eq:Rabi_oscillations}
H^{\prime} = \frac{\delta}{2} \sigma_z + \frac{\Omega_{\rm t} (1 + \epsilon_{\rm t})}{2}\sigma_x,
\end{equation}
where we assumed $\xi=0$ without loss of generality.  The detuning $\delta$ generates an additional rotation about the $z$-axis of the Bloch sphere and the amplitude error $\epsilon_{\rm t}$ shifts the rotation speed around the $x$-axis. Spatial and temporal variations of $\delta$ and $\epsilon_{\rm t}$ due to noise create different dynamics within the ensemble and result in a decay in the experimentally observed Rabi oscillations when averaged \cite{de_raedt_quantum_2012}. The decay rate of Rabi oscillations increases when the variation of the detuning $\delta$ is large in comparison to $\Omega_{\rm t}$, as well as for higher relative amplitude errors $\epsilon_{\rm t}$. In the context of Rabi magnetometry, this behavior sets a lower bound of the measurable amplitude range. 

CCDD overcomes this limitation and can be used for sensing of oscillating magnetic fields. 
The energy diagram of the system under CCDD is shown in Fig.~\ref{fig_diagram}(a). Given a two-level system (TLS), 
the first drive creates dressed states with an ideal frequency separation  
$\Omega_1\gg\delta$ to ensure robustness \cite{cai_robust_2012,genov_mixed_2019}. This is evident when we consider the energy gap in the first dressed basis in the absence of other fields $\sqrt{\Omega_1^2+\delta^2}\approx \Omega_1+\delta\frac{\delta}{2\Omega_1}$, where the last approximation is valid for $\delta\ll\Omega_1$. Thus, the effect of $\delta$ is suppressed by a factor of $\delta/(2\Omega_1)$, compared to the bare basis. However, the amplitude error $\epsilon_1$  
shifts the frequency separation to $\Omega_1\to\Omega_1(1+\epsilon_1)$, leading to dephasing in the dressed basis. To overcome this effect, a second drive is applied resonantly with the first drive. Likewise, when 
$\Omega_2$ is greater than the frequency shift induced by the first drive error $\Omega_1\epsilon_1$  \cite{cai_robust_2012,genov_mixed_2019}, the system gains further robustness. 

Several versions of CCDD are possible, e.g., with \cite{genov_mixed_2019,salhov_protecting_2024} or without \cite{stark_narrow-bandwidth_2017, fang_quantum_2024} initial and readout $\pi/2$ pulses 
after preparing the system in the ground state $|0\rangle$.
We adopt the former scheme, which we label transverse CCDD, because it allows for spin locking with the first driving field, so the signal oscillates only at the slow frequency $\Omega_2$ in the absence of a target field \cite{genov_mixed_2019,salhov_protecting_2024}. Figure~\ref{fig_diagram}(b) shows the pulse sequence used for CCDD magnetometry. After initial preparation in state $|0\rangle$ by a green laser, a $\pi_y / 2$ pulse prepares a coherent superposition along the $x$-axis of the Bloch sphere in the bare basis. The target signal is detected during CCDD with two driving fields. After a final $\pi_{ y}/2$ ($\pi_{-y}/2$) pulse, the population of the x state is mapped onto the populations of state $|0\rangle$ ($|1\rangle$). The latter are readout by detecting state-dependent red-shifted photoluminescence excited by the green laser.  
The differential signal from the measurements with the alternating phase of the last $\pi/2$ pulse suppresses common-mode noise, primarily originating from laser fluctuations, and is referred to as the signal throughout this article.

Finally, we detect the amplitude of a target MW signal when its frequency $\omega_{\rm t}$ is resonant with the transition frequency between the dressed states in the doubly-dressed basis  
\cite{cai_robust_2012,stark_clock_2018,genov_mixed_2019,salhov_protecting_2024}.  
In previous work \cite{stark_narrow-bandwidth_2017}, the condition was $\omega_{\rm t} = \omega_0 - \Omega_1 - \Omega_2$. Recently, Salhov {\it et al.} demonstrated improved sensitivity under a low-attenuation condition $\omega_{\rm t} = \omega_0 - \Omega_2$ \cite{salhov_protecting_2024}. In this work, we adopt the latter approach, which also improves robustness \cite{stark_narrow-bandwidth_2017,genov_mixed_2019}.

\section{CCDD in the presence of inhomogeneities}\label{sec:CCDD}

\subsection{Experimental implementation}
We perform our experiments with an ensemble of negatively-charged nitrogen-vacancy (NV) centers in diamond  \cite{reddy_two-laser_1987,oort_optically_1988, doherty_nitrogen-vacancy_2013}. The experimental conditions relevant to the inhomogeneities of the fields can be summarized as follows with further experimental details given in 
Appendix~\ref{sec:experiment}.
Figure~\ref{fig_diagram}(c) shows a schematic of the experimental setup  around the diamond sample. Specifically, we use a single crystal diamond containing a large number of NV center, which is irradiated with green laser light for initialization and readout \cite{jelezko_observation_2004}. The diamond sample was created by the high-pressure high-temperature (HPHT) method from $^{12}$C-enriched solid carbon source, followed by electron beam irradiation, resulting in 99.995 \% $^{12}$C abundance for reduced spin noises. The NV$^-$ and P1 concentrations are $\approx$ 0.1 ppm and 1.6 ppm, respectively.
The NV centers contributing to the measurement signal are roughly defined by the overlap between the NV-doped diamond and the laser excitation profile. The sample is 1 mm $\times$ 1 mm in area and 500 $\mu$m thick. 
The laser is focused down to a diameter of 98 $\mu$m. The effective measurement volume can be approximated by a cylinder with a diameter of 98 $\mu$m and a height of 500 $\mu$m. We consider inhomogeneities in detuning and amplitude noise across this volume. 

A bias magnetic field is applied using a pair of neodymium magnets and aligned along one of the NV axes. 
We perform coherent control of the NV electron spins by a planar MW antenna placed next to the diamond \cite{sasaki_broadband_2016}. 
Owing to the near field nature of the microwave delivery and the finite spatial extent of the NV ensemble, the antenna produces a spatially varying field amplitude across the sensing volume (see Ref.~\cite{rezinkin_uniform_2024} for a comparison of field inhomogeneity away from the antenna plane with barrel-shaped MW forming systems.)
In this ensemble, the longitudinal relaxation time $T_1$ and spin coherence time $T_2$ are measured to be 5 ms and 47 $\mu$s, respectively (see Appendix~\ref{sec:characteristics}). 
Importantly, the dephasing time  $T_2^*$ is 701 ns, characterizing the inhomogeneity in detuning.

\begin{figure}[t!]
\includegraphics[width=8.6cm]{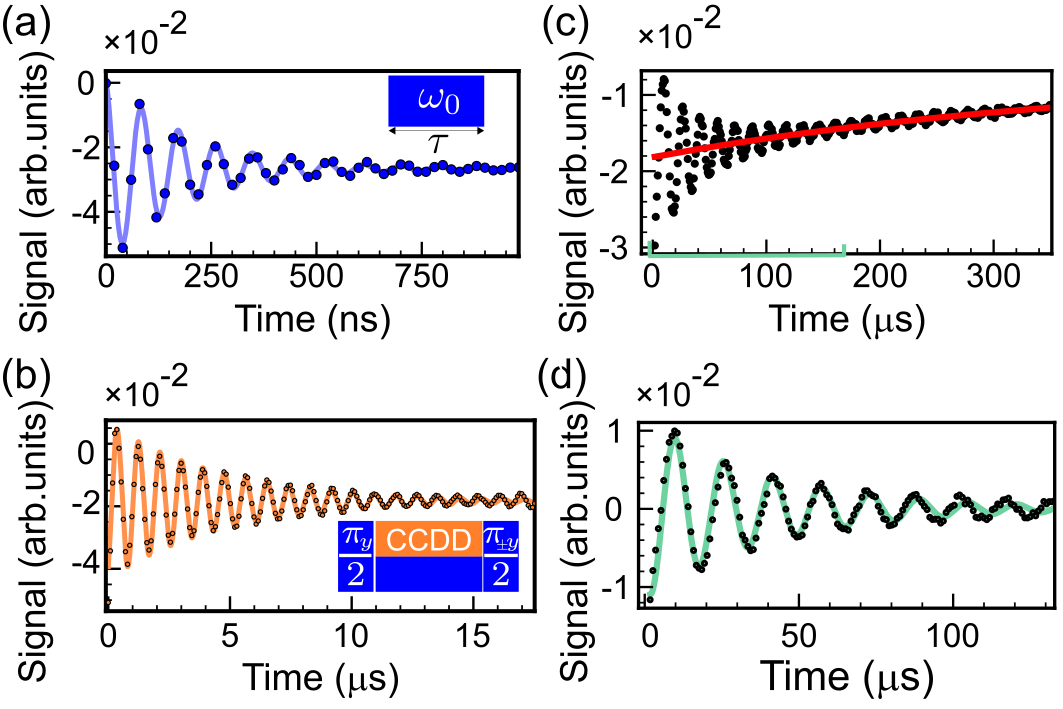}
\caption{\label{fig_characterization} (a) Rabi oscillations under the single drive.  A microwave field is applied resonantly with the central line $\omega_0 =  (2\pi)\,$2.7081 GHz of the hyperfine sublevels from the $^{14}$N. The rapid decay of the Rabi oscillations reflects the inhomogeneity in the drive amplitude across the ensemble. The Rabi frequency of the first drive $\Omega_1$ is kept $(2\pi)\,$11.3 MHz throughout this article. (b) Rabi oscillations at angular frequency $\Omega_2\approx \Omega_1 / 10$ under the transverse CCDD sequence. They are sampled at time steps of $\tau_{\Omega_1} = 2\pi/ \Omega_1$ when the first dressed basis corresponds to the bare basis in the absence of noise.  
(c) Dynamics of the system under CCDD with a test MW signal. $\Omega_1=(2\pi)\,$11.3 MHz and $\Omega_2=(2\pi)\,$1.13 MHz are used for the CCDD, and the frequency of the target signal is set to $\omega_{\rm t} = \omega_0 - \Omega_2=(2\pi)\,$2.7070 GHz. The time increment is set to $\tau_{\Omega_2} = 2\pi/\Omega_2$. The background curve is fitted with a single-exponential function, which also appears in CCDD without the target signal 
(see Appendix~\ref{sec:CCDD_CP} for details.)
(d) Initial oscillations from the same measurement after background subtraction. The data is well fitted with an exponentially decaying sinusoidal function, yielding $\Omega_{\rm t}^{\prime}$ = $(2\pi)\,63.2$ kHz.
}
\end{figure}

We first determine the Rabi frequency $\Omega_1$ by driving the system with a single resonant driving field.  The resulting Rabi frequency is $\Omega_1\approx (2\pi)\,11.3$ MHz, as shown in \ref{fig_characterization}(a). The amplitude inhomogeneity is estimated from the decay of Rabi oscillations, which is $\sim \exp\{-(\Omega_1 t\sigma_{\epsilon })^2/2\}$ when we apply a strong field, where $ \sigma_{\epsilon }\approx 0.1$ is the standard deviation of $\epsilon_k$, which we assume to have a normal distribution \cite{de_raedt_quantum_2012}
(see Appendix~\ref{sec:characteristics} for details).

\subsection{CCDD}
We apply transverse CCDD by setting the carrier frequencies of the driving MW fields to $\omega_0$. In addition, the modulation frequency of the second field is set  
$\widetilde{\Omega}_1=\Omega_1$.
We note that shifting $\widetilde{\Omega}_1$ from $\Omega_1$ could improve the coherence times further if the amplitude noise is correlated \cite{salhov_protecting_2024}. 
We use broadband composite pulses BB1 instead of simple rectangular $\pi/2$ pulses to improve the initialization fidelity 
(see Appendices~\ref{sec:CCDD_rect} and \ref{sec:CCDD_CP} for comparisons).
Figure~\ref{fig_characterization}(b) shows the result of the CCDD sequence when the input voltage for the second drive is set on a fifth of that of the first drive $V_{\Omega_2} = V_{\Omega_1} /5 = 30$ mV, leading to $\Omega_2/\Omega_1\approx 0.1$. By fitting the observed oscillations with an exponentially decaying cosine function, we extract $\Omega_2\approx(2\pi)\,$1.13 MHz. The decay time is prolonged to $T_{\Omega_2}$ = 4.6 $\mu$s in comparison to single drive decay $T_{\Omega_1}= $ 197 ns mainly due to the suppression of the amplitude noise of the first drive. 

 We calibrate the CCDD period $\tau_{\Omega_2}=2\pi/\Omega_2$ 
to be an integer multiple of the first drive Rabi period by adjusting the second drive amplitude. This allows us to isolate the modulation induced by the target signal by setting the time step as integer multiples of the CCDD period. As discussed in Appendix~\ref{sec:CCDD_rect}, we
observe an exponential decay background in the measured signal dependent on the excitation profile of the $\pi/2$ pulse, which we attribute to the hyperfine side bands. 

\section{Magnetometry with CCDD}\label{sec:magnetometry}
We demonstrate GHz range magnetometry by applying a test target signal in addition to the CCDD sequence. Here, we focus on measuring the amplitude of the target signal when its frequency is set as $\omega_{\rm t} = \omega_0 - \Omega_2$ = $(2\pi)$ 2.7070 GHz, as proposed in \cite{salhov_protecting_2024}.

\begin{figure*}[t]
\includegraphics[width=17.2cm]{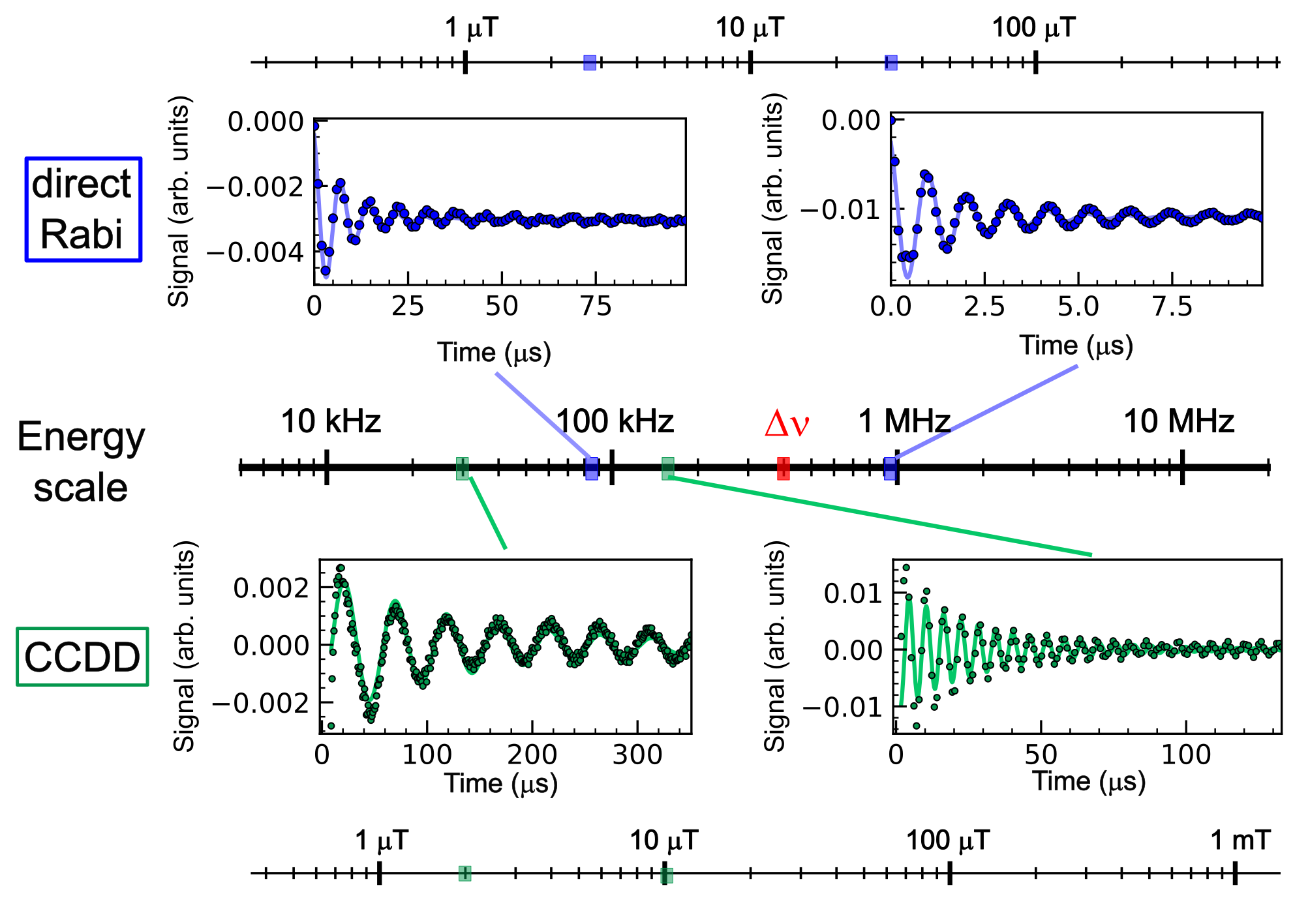}
\caption{\label{fig_comparison} Dynamics of the system under direct Rabi (top) and CCDD (botttom) with varying amplitudes of the target MW signal, shown in comparison to the system linewidth $\Delta\nu$ = 415 kHz. When the amplitude of the target MW signal is strong, such that its Rabi frequency exceeds the linewidth, the signal drives the system effectively,  allowing for robust measurement (top right). See Appendix~\ref{sec:magnetometry} for details on the amplitude dependence. In contrast, when the signal amplitude is smaller than the linewidth, the Rabi oscillations suffer from the detuning, leading to rapid decrease in the measurement contrast (top left). This behavior determines a lower bound of the measurable amplitude of target MW signal. The top axis converts the Rabi frequency into the corresponding magnetic field amplitude (in tesla). In the CCDD case (bottom), the dressed state created by the CCDD enables robust detection of weaker signals (bottom right). As a result, CCDD extends the minimum detectable amplitude (bottom left). The bottom axis converts the oscillation frequency into the corresponding magnetic field amplitude (in tesla).}
\end{figure*}

Figure~\ref{fig_characterization}(c)
shows the system dynamics when the amplitude of the target signal is set to $V_{\rm t}$ = 2 mV 
(see Appendix~\ref{sec:CCDD_sig}
for the results with different amplitudes). The interaction time $\tau$ is chosen as integer multiples of the second oscillation period $\tau_{\Omega_2} = 10 \tau_{\Omega_1}$. Note that the measurement signal is obtained by subtracting the signals from the alternating measurements with $\pm y$ phases. 
Exponentially decaying sinusoidal oscillations are observed on top of a single-exponential decay, indicated by the red curve. We attribute this background to the limited control over the sidebands by monitoring the dynamics with different drive strengths as well as with different excitation $\pi/2$ pulses using composite pulses 
(see Appendices~\ref{sec:CCDD_rect}-\ref{sec:spin_lock}).

Figure~\ref{fig_characterization}(d)
shows a zoom-in view of the data after removing the exponential background. The initial oscillations fit well to an exponentially decaying sinusoidal functions. The extracted frequency is 63.2 kHz, which agrees with the theoretical expectation $\Omega_{\rm t}^\prime = \frac{1}{2}\Omega_{\rm t}$ under this condition \cite{salhov_protecting_2024}. 
The coherence time is extended to $T_{\Omega_{\rm t}}$ = 40.4 $\mu$s owing to suppression of the second-drive amplitude noise by the target signal.
The contrast of the oscillations is reduced by more than a factor of 2 compared to the Rabi oscillations driven by a strong Rabi frequency of $(2\pi)\,$11.3 MHz in  Fig.~\ref{fig_characterization}(a). This reduction reflects both the decaying non-oscillatory background component and the reduced contrast of oscillations. Specifically, the transient dynamics of an ensemble system under transverse CCDD depend strongly on the excitation profile and the amplitude of the drive fields. 
A subgroup of the ensemble is controlled with limited efficiency by CCDD and contribute to the non-oscillatory background. We consider that the primary source of this imperfect control originates from the hyperfine detuning of the sidebands from the previous analysis in combination with a limited strength of the first driving field.  
While CCDD with a strong drive ($\Omega_1 \gg A_{\parallel} > \Delta\nu$) firmly dresses the central line, the sidebands are only partially dressed due to their effective detuning of several MHz. As a result, only the resonantly dressed central line contributes effectively to the detection of weak target MW signals. Nevertheless, the observed improvement in coherence time with CCDD suggests the possibility of detecting weaker signals.

\subsection{Detection range comparison}
To examine the range of possible amplitudes to be measured, we vary amplitudes of the test target signal and compare CCCD with the direct Rabi method, which is the current state-of-the-art technique for GHz signal sensing with NV ensembles \cite{alsid_solid-state_2023}. Figure~\ref{fig_comparison} maps the observed dynamics for both methods 
, while a detailed amplitude comparison is provided in Appendix~\ref{sec:CCDD_sig}.
The central energy scale indicates oscillation frequencies relative to the linewidth $\Delta\nu$ = 415 kHz.  
The upper portion of Fig.~\ref{fig_comparison} shows the general behavior of the direct Rabi method. The top axis represents the corresponding Rabi frequency $\Omega$, expressed in terms of magnetic field amplitude via $B_{\rm} = \Omega / \gamma$. As shown in the upper right panel, the direct Rabi method performs well when the signal amplitude is much greater than the linewidth $\Delta\nu$. In contrast, the upper left panel shows the transient dynamics when the amplitude is smaller than the linewidth $\Delta\nu$, where Rabi oscillations are strongly damped. This behavior can be qualitatively explained as follows:  
when the signal amplitude is smaller than $\Delta\nu$, the detuning damps the Rabi oscillations, setting a lower bound on the measurable amplitude in GHz-range AC magnetometry.  
When the amplitude exceeds the hyperfine splitting, the sidebands are also excited, and special care must be taken in analyzing the resulting dynamics \cite{fedder_towards_2011}. 

On the other hand, the data in the lower half of Fig.~\ref{fig_comparison} show the general behavior of CCDD magnetometry. The bottom axis represents the oscillation frequency $\Omega_{\rm t}^{\prime}$, expressed in terms of the magnetic field amplitude $B_{\rm t} = \frac{\Omega_{\rm t}^{\prime}}{2\gamma}$. As shown in the bottom right of Fig.~\ref{fig_comparison}, the decay of the oscillations is more gradual compared to the Rabi measurements at similar amplitudes. We note that amplitude inhomogeneity in the target field can also limit the coherence time with CCCD but its effect is small for very weak target signals, compared to the one of the driving fields. Remarkably, the data in the bottom left shows that CCDD method can detect weak signals as low as 25.9 kHz. Although the contrast again decreases because the target signal cannot drive the doubly dressed state robustly due to the hyperfine sidebands and the large amplitude inhomogeneity, the prolonged coherence time enables the detection of much weaker amplitudes of the target signal. 

\subsection{Sensitivity}\label{ssec:sensitivity}

\begin{figure}[t]   
\includegraphics[width=8.6cm]{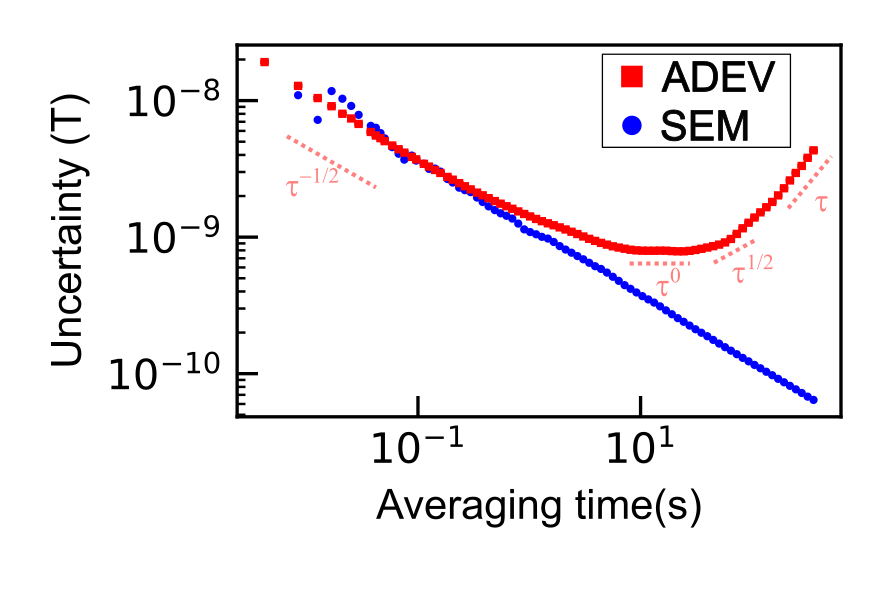}
\caption{\label{fig_uncertainty}  Uncertainty evaluation. The amplitude of the target micrwave (MW) $B_{\rm t}$ are repeatedly measured under CCDD with the test MW signal. The amplitude is the same as in Fig.~\ref{fig_characterization} and the interaction time is kept constant at $\tau$ = 67 $\mu$s. The Allan deviation (ADEV) and the standard error of the mean (SEM) for the measured magnetic field amplitude $B_{\rm t}$ are shown in red squares and blue circles, respectively. Pink dashed lines are guides to the eye for the corresponding $\tau$-scalings. The sensitivity of this amplitude measurement is calculated to be 956 pT/$\sqrt{\rm Hz}$, based on the slope of the SEM 
(see Appendix~\ref{sec:magnetometry_details}.)
}
\end{figure}

In the following, we calculate the sensitivity in our proof-of-principle experiment for measuring the amplitude of a GHz oscillating field with an NV ensemble with large inhomogeneities
, with the detailed calibration procedure described in Appendix~\ref{sec:magnetometry_details}.
We measure the signal variation when we 
sweep the amplitude of the test signal,
while keeping the interaction time $\tau$ fixed at 67 $\mu$s 
at a node of the responsivity curve for slope detection \cite{degen_quantum_2017}. Then, the response is approximately linear to the amplitude, 
and its inverse function is used to obtain the 
magnetic field amplitude 
$B_{\rm t}$ from the measurement signal $S$. 

For test purposes, we fix the input voltage at the node and measure the amplitude of the test MW magnetic field. This measurement is repeated to reduce the measurement uncertainty. 
We calculate the Allan deviation to characterize the measurement stability \cite{allan_should_1987, zhang_allan_2008, benkler_relation_2015}. Figure~\ref{fig_uncertainty} shows the Allan deviation (red squares) and the standard error of mean of this measurement (blue circles). 
The Allan deviation scales as $\sim \tau^{-1/2}$ up to $\tau$ = 0.1 s and then gradually deviates through an intermidiate crossover regime toward a plateu, followed by an upward bending at longer averaging times. The initial $\tau^{-1/2}$ scaling indicates the white noise, providing the magnetic sensitivity 956 pT / $\sqrt{{\rm Hz}}$. The subsequent deviation shows the accumulation of colored noise. In particular, the plateau $\tau^0$ indicates the flicker frequency noise, while the upward bending at the long averaging times is indicative of long-term drift. A detailed quantitative decomposition of the underlying noise sources is beyond the scope of the present work.  
We calculate the uncertainty of the amplitude measurement by the standard error of mean within the white-noise regime. Using data acquired over 0.4 s, the amplitude of the test signal is measured as $B_{\rm t} = 2197 \pm 96$ nT 
(see Sec.~\ref{ssec:uncertainty} for detailed discussion on the white-noise regime).

\section{Discussion}\label{sec:discussion}

In this work, we demonstrated GHz-range AC magnetometry by applying concatenated continuous dynamical decoupling (CCDD) to an ensemble of NV centers in diamond under inhomogeneous drive fields. We observed unique transient dynamics of the system driven by a test target signal under CCDD, which are mainly due to the present hyperfine sidebands and the large inhomogeneity of control fields. We then demonstrated that CCDD enables the measurement of weaker signal amplitudes by comparing the transient dynamics under CCDD to those observed with the direct Rabi method across a range of test MW signal amplitudes, relative to the system linewidth $\Delta\nu$. These results provide practical guidance for implementing GHz-range AC magnetometry in ensemble systems subject to detuning of the system and inhomogeneity in the target signal amplitude. 

The general behavior of transient dynamics under the direct Rabi method is consistent with the results of Alsid {\it et al.} \cite{alsid_solid-state_2023} in that the signal deteriorates when the Rabi frequency is smaller than the system linewidth. They also reported that the sensitivity is optimized when the Rabi frequency slightly exceeds the linewidth ($(2\pi)$35 kHz in their system). 
This effect
was not observed in our measurement, most likely due to a shift in the zero-field splitting of the NV centers resulting from the heating by laser irradiation. Nevertheless, our results demonstrate that even weaker signals could be measured using CCDD method, even in the presence of significant linewidth broadening and driving field inhomogeneity, which are typical for spin ensembles. The advantage of CCDD method becomes especially prominent when detunings are large compared to the signal amplitude. This approach could provide a promising solution for systems with a broadened spectral linewidth or limited driving field strength. 

We note previous independent work demonstrating CCDD magnetometry with an ensemble of boron vacancy centers in hBN \cite{patrickson_high_2024}, where large detunings of approximately 150 MHz dominate over amplitude noise. In that study, a strong first drive of 100 MHz was applied to dress as many spins as possible within the broadened energy spectrum. Our results on hyperfine lines may be relevant if the energy spectrum contains distinguishable hyperfine lines through isotope engineering \cite{sasaki_nitrogen_2023}. In such cases, it is no longer clear whether applying a strong first drive remains optimal. Under these conditions, the coherence time of the dressed states would ultimately be limited by the short relaxation time of hBN electron spins of $T_1 < $ 20 $\mu$s, which is about three orders of magnitude shorter than with NV electron spin ensembles and severely limits the detectable amplitude. Further studies are necessary to examine the optimal driving field amplitudes across different systems with different varying linewidths and hyperfine sublevels.

The sensitivity reported here reflects the operating conditions adopted in this work. To prevent saturation in the photodetector under pulsed operation, the excitation laser power is limited, resulting in an initialization time of 2 ms. We additionally observe the spectral shift in the energy spectrum attributed to the laser-induced heating. Crucially, the demonstrated sensing protocol remains robust under these non-ideal yet practically relevant conditions as in miniaturized implementations \cite{sturner_compact_2019, sturner_integrated_2021}.

Our study provides important insights into the current research on Rabi-based AC magnetometry using defect qubits. The frequency bandwidth of such approaches is primarily determined by the energy level of the system. In the case of diamond NV centers, it is given by the 2.87 GHz zero-field splitting and the Zeeman shift. 
Our protocol is applicable for wide frequency ranges by tuning the bias magnetic field.
For lower transition frequency, 
the first drive must be set much lower than the energy gap to satisfy the rotating wave approximation, and the parallel component of the control field may no longer be negligible \cite{yudilevich_coherent_2023,deshmukh_observation_2025}. Alternatively, different types of defect qubits with different zero-field splittings could be employed to target other frequency ranges. These findings could also be applicable to current research on Rabi-based magnetometry with defect qubits in applications such as microwave device characterization \cite{shao_diamond_2016,yang_noninvasive_2018, chen_vectorial_2020,chen_microwave_2024,jia_near-field_2024, robertson_radiofrequency_2025, zeng_towards_2025}, spin wave detection \cite{van_der_sar_nanometre-scale_2015, bertelli_magnetic_2020, ogawa_wideband_2025}, high-field nano- or micro-scale NMR \cite{hermann_extending_2024, london_detecting_2013, maier_efficient_2025}, among others. CCDD is complementary can in principle be combined with  heterodyne detection for improved frequency resolution, e.g., similarly to \cite{staudenmaier_phase-sensitive_2021} but this goes beyond the scope of this work.

\section{Conclusion}\label{sec:conclusion}
In summary, this study demonstrated GHz-range AC magnetometry using a large ensemble of NV centers under inhomogeneous drive fields, enabled by concatenated continuous dynamical decoupling (CCDD). CCDD robustly dressed the ensemble spins and allowed for the detection of weaker magnetic fields. 

Beyond solid-state defect qubits, this technique can be readily applied to a wide range of quantum systems. The CCDD method is compatible with many ongoing studies of Rabi-based AC magnetometry. Its application to spin defects with higher zero-field splittings will extend the detection bandwidth and broaden the utility of solid-state quantum sensors.

\begin{acknowledgments}
The work was supported by the German Federal Ministry of Research, Technology and Space (BMFTR) via future cluster QSENS and projects EXTRASENS (13N16935) DIAQNOS (No. 13N16463), quNV2.0 (No. 13N16707, DLR via project QUASIMODO (No. 50WM2170), Deutsche Forschungsgemeinschaft (DFG) via projects 386028944, 387073854, 445243414, 491245864, 499424854, 532771161, 546850640, and joint DFG/JST ASPIRE program via project 554644981, European Union's HORIZON Europe program via projects QuMicro (No. 101046911), SPINUS (No. 101135699), CQuENS (No. 101135359), QCIRCLE (No. 101059999) and FLORIN (No. 101086142), European Research Council (ERC) via Synergy grant HyperQ (No. 856432), IQST and Carl-Zeiss-Stiftung. 
A. S. gratefully acknowledges the support of the Clore Israel Foundation Scholars Programme, the Israeli Council for Higher Education, and the Milner Foundation. A. R. acknowledges the support of ERC grant QRES, project number 770929, Quantera grant MfQDS, the Israeli Science foundation (ISF), the Schwartzmann university chair and the Israeli Innovation Authority under the project ``Quantum Computing Infrastructures''. S.O. gratefully acknowledges support from the Japan Science and Technology Agency (JST) via ASPIRE program (No. JPMJAP24C1).
\end{acknowledgments}

\appendix

\section{Experimental details}\label{sec:experiment}
\subsection{Setup}
\begin{figure}[t]
\includegraphics[width=8.6cm]{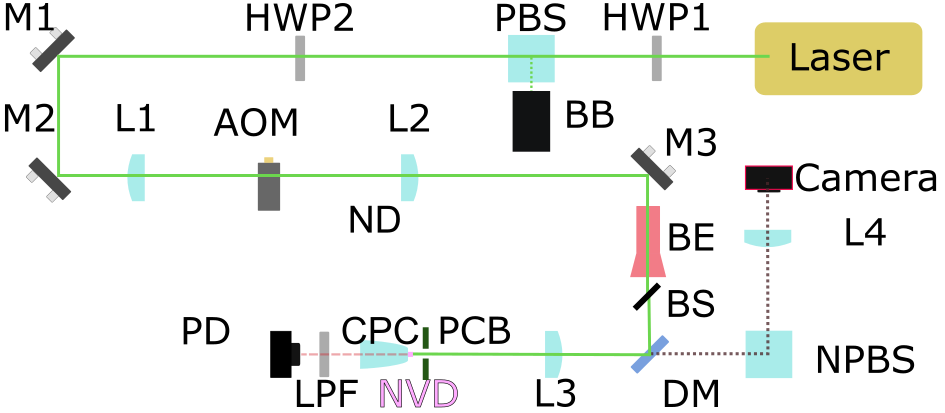}
\caption{\label{fig:schematic} Schematic of the optical setup. HWP: half-wave plate; PBS: polarising beam splitter; BB: beam blocker; M: mirror; L: lens; AOM: acousto-optic modulator; BE: beam expander; DM: dichroic mirror; NPBS: non-polarizing beam splitter; PCB: printed circuit board; NVD: NV containing diamond; CPC: compound parabolic concentrator; LPF: long pass filter; PD: photo detector.}
\end{figure}

\paragraph{Optics}
The optical setup in our experiment is schematically shown in Fig.~\ref{fig:schematic}. The NV centers are optically excited by a green laser, which is pulsed by an acousto-optic modulator (AOM). A continuous-wave green laser with a wavelength of 532 nm is output from a diode-pumped solid-state laser (Spectra Physics, Millennia eV5). The laser power can be tuned by a pair of a half-wave plate (Thorlabs, WPH10M-532) and a polarizing beam splitter (Thorlabs, PBS251); the rest of the beam is blocked with a beam blocker (Thorlabs, BT620). The beam is focused onto the AOM (Crystal Technology 3250-220) using a plano-convex lens (Thorlabs, AC-254-300-A-ML), after adjusting its polarization with another half-wave plate (Thorlabs, WPH10M-532). The first-order diffraction is collimated by another plano-convex lens (Thorlabs, AC-254-300-A-ML), and its diameter is adjusted using a beam expander (Thorlabs, ZBE22). This beam is then focused on the diamond with a diameter of approximately 98 $\mu$m. The red-shifted photoluminescence (PL) from the NV centers is collected using a compound parabolic concentrator (CPC) (Edmund, \#17-709), which is attached to the diamond using optical glue (M-GLASS). The collected PL passes through a long-pass filter (AHF Analysentechnik, BLP01-633R-25) and is detected with a photodetector (Thorlabs, PDA100A2). Finally, the analog output is digitized by an analog-to-digital converter (ADC) (Spectrum Instrumentation, M4i.4420-x8) and processed by a computer. The beam diameter is checked by a camera placed behind a dichroic mirror (Thorlabs, DMLP650). The photoluminescence is collected by a lens and imaged by another lens onto a camera (Thorlabs, CS165MU).

\paragraph{MW Electronics}
All MW signals are generated by an arbitrary waveform generator (Keysight, M8915A), including the first drive, the second drive, and the test target signal. After amplification by a MW amplifier (ar 200S1G4A), the signals are fed into a planar antenna \cite{sasaki_broadband_2016}. 

\paragraph{Sample}
The diamond sample was created by the high-pressure high-temperature (HPHT) method, followed by electron beam irradiation and annealing. A $^{12}$C-enriched diamond crystal was grown by the temperature gradient method at HPHT conditions of 5.5 GPa and 1350 $^\circ\mathrm{C}$ using a Fe-Co-Ti solvent and a $^{12}$C-enriched solid carbon source obtained by pyrolysis of 99.999 \% $^{12}$C-enriched methane. The crystal was irradiated with 2 MeV electrons to a fluence of $1\times10^{17}\mathrm{cm}^{-2}$ at room temperature and subsequently annealed at 1000 $^\circ\mathrm{C}$ for 2 hours in vacuum. 
The $^{12}$C isotope ratio was measured to be 99.995 \% by secondary ion mass spectrometry (SIMS). A (100)-oriented plate with dimensions of  $1\times1\times0.5$ mm$^3$ was obtained from the (100) growth sector of the crystal by laser-cutting, followed by polishing.

\paragraph{Pulsed measurement}
The initialization time for the ensemble of NV centers was determined by measuring the PL intensity under laser illumination after inverting the spin state by a $\pi$ pulse \cite{nizovtsev_quantum_2005}. For each measurement, the sample was irradiated with green laser light for 2 ms. The measurement signal used for spin state estimation was obtained by the average PL intensity in the initial 10 $\mu$s divided by the average PL in the final 10 $\mu$s, to mitigate the variation in the input laser power after the AOM. This was accomplished by gating the ADC during the corresponding time widnows. All the experiments are controlled by our custom measurement software \cite{binder_qudi_2017}.

\subsection{Beam waist}
\begin{figure}[h]   
\includegraphics[width=4cm]{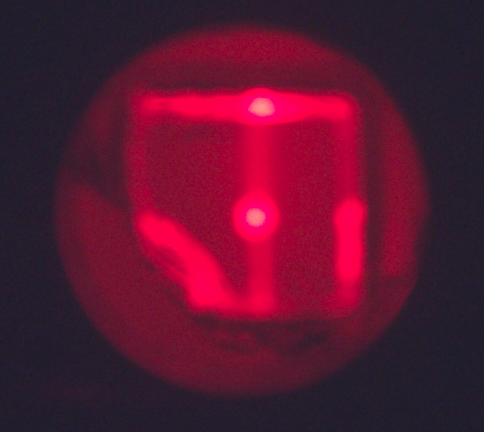}
\caption{\label{fig:image} Focus check. The beam diameter is measured to be 98 $\mu$m.}
\end{figure}

The beam diameter of the irradiated green laser is estimated from the image of photoluminescence (PL) from the diamond, as captured by the camera shown in Fig.~\ref{fig:image}. The diamond is viewed through a circular hole in the PCB board. The PL is collected by a lens placed in front of the diamond and imaged onto the camera after passing through a dichroic mirror. The physical length of the diamond was measured to be 950 $\mu$m using a commercial microscope, and this value is used to calibrate the image scale. The full width at half maximum (FWHM) of the beam is estimated to be 98 $\mu$m.

\subsection{Initialization time}
\begin{figure}[h]   
\includegraphics[width=8.6cm]{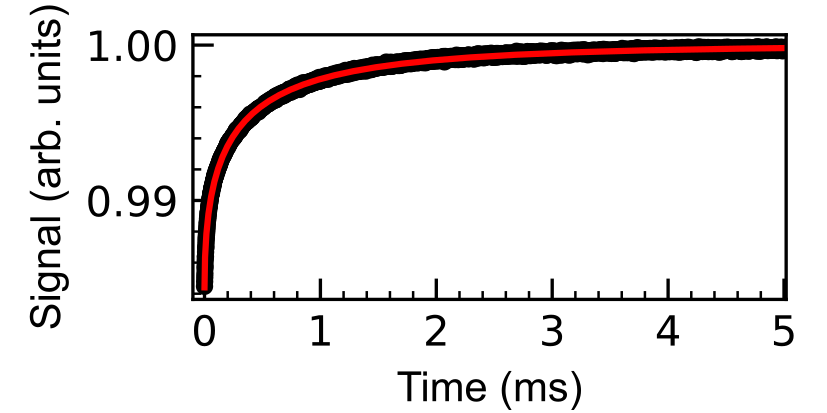}
\caption{\label{fig:init_time} Initialization curve. The red curve is a fit to a stretched exponential: $A\exp[-(t/T)^p]+C$, where $p$ = 0.50 and $T$ = 258 $\mu$s.}
\end{figure}

The initialization time for the measurement is determined by the method in \cite{nizovtsev_quantum_2005}. Green laser is applied and MW is applied on resonant with the system. After waiting until it reaches its steady state, the MW is switched off. Figure~\ref{fig:init_time} shows the amount of PL when the green laser is applied continuously after the switch off. Increase of the PL is observed because the ensemble NV centers are initialized into $m_{s}$ = 0 state. $t_{\rm I}$ = 2 ms is used as initialization for the measurements in the main.

\subsection{Note on transverse CCDD}
As discussed in the main text, adoption of transverse CCDD is beneficial for improved response of the system in magnetometry. In addition, we consider that this approach is suitable when the subtraction of two alternating-phase measurements is adopted to remove the common-mode background noise \cite{wolf_subpicotesla_2015}. This is particularly relevant for a large ensemble of NV centers because colored background noise can arise, for example, due to the time-dependent performance of an acousto-optic modulator (AOM) \cite{wolf_subpicotesla_2015}. Longitudinal CCDD can be implemented by applying inversion pulses at the end of an alternating measurement, but this introduces a temporal offset between the alternating measurements, potentially increasing differential-mode noise due to variations in the input laser power. This effect can be suppressed with transverse CCDD by shifting the phase of $\pi/2$ pulses.

\section{System characteristics}\label{sec:characteristics}
This 
Appendix summarizes the characteristics of the ensembel system, including the energy spectrum, the inhomogeneities in the detuning and drive-field amplitude, and the characteristic relaxation time ($T_1, T_2^*$, and $T_2$).

\subsection{Energy sperctrum}

Fig.~\ref{fig:characterization}(a) shows the energy spectrum of the system using pulsed optically detected magnetic resonance (ODMR). Three dips are observed as a result of the hyperfine interaction with the $^{14}$N spin ($I$=1) \cite{felton_hyperfine_2009}, corresponding to $A_{\parallel}$ = 2.16 MHz, in addition to the central line at $\omega_0$ = $(2\pi)$2.7081 GHz. The inhomogeneity in the detuning is quantified by the linewidths of these dips \cite{barry_sensitivity_2020}. A Lorentzian fit yields a linewidth $\Delta \nu$ = 415 kHz. 

\subsection{Drive inhomogeneity}

\begin{figure}[t]
\includegraphics[width=8.6cm]{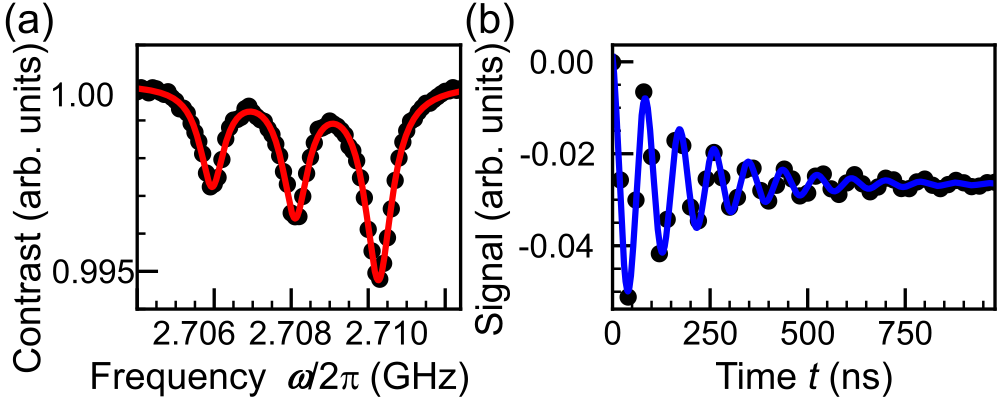}
\caption{\label{fig:characterization} (a) Energy of the system obtained by pulsed optically detected magnetic resonance (ODMR) of an ensemble of NV centers. In addition to the central line at $\omega_0$ = $(2\pi)$2.7081 GHz, sidebands are observed as a result of the hyperfine interaction with $^{14}$N nuclear spins ($I$=1). (b) Rabi oscillations with a MW drive field resonant with the central frequency $\omega_0$. Fitting with the function $A_{\Omega_1}\exp[-(\tau/T_{\Omega_1})]\cos[\Omega_1 t] + C_{\Omega_1}$ yields a Rabi frequency $\Omega_1$ = $(2\pi)$ 11.36 MHz and decay time $T_{\Omega_1}$ = 197 ns. }
\end{figure}

The inhomogeneity of the drive amplitude is characterized by the decay of Rabi oscillations \cite{de_raedt_quantum_2012}. Fig.~\ref{fig:characterization}(b) shows the Rabi oscillations when the MW field is resonant with the central line of the spectrum. Oscillations decay rapidly, which originates from the inhomogeneity in the drive amplitude. Fitting the data with $A_{\Omega_1}\exp[-(\tau/T_{\Omega_1})]\cos[\Omega_1 t]$ yields $\Omega_1$ = $(2\pi)$11.36 MHz, $A_{\Omega_1}$ = 0.0287 and $T_{\Omega_1}$ = 197 ns, which corresponds to $\approx 2.24$ periods of the Rabi oscillation. 

Assuming a Gaussian distribution of the relative inhomogeneity error $\epsilon_1\sim N(0,\sigma_\epsilon)$, one can show that the probability that the NV center stays in state $|0\rangle$ during Rabi oscillations by a strong single driving field is  
\begin{equation}
P_{|0\rangle}=\frac{1}{2} \left(1+\cos (\Theta) e^{-(\Theta\sigma
   _{\epsilon })^2/2}\right),
\end{equation}
where $\Theta=\Omega_1 t$ is the pulse area of the single driving field during the Rabi oscillations and we have neglected the effect of detuning errors and the hyperfine interaction due to the strong driving field. In order to obtain $T_{\Omega_1}$, we solve $(\Omega_1 T_{\Omega_1}\sigma
   _{\epsilon })^2/2=1$. Thus, we obtain that $T_{\Omega_1}=T_{\text{Rabi}}/(\sqrt{2} \pi\sigma_\epsilon)=\sqrt{2}/(\Omega_1\sigma_\epsilon)$, where $T_{\text{Rabi}}=2\pi/\Omega_1$ is the Rabi oscillation period, so we estimate that $\sigma_\epsilon\approx 0.1$ in our particular experiment. 

\subsection{Longitudinal relaxation $T_1$}
\begin{figure}[h]
\includegraphics[width=6cm]{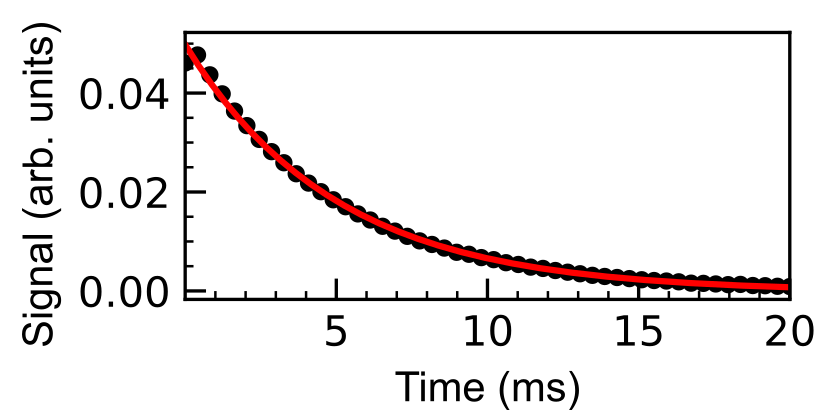}
\caption{\label{fig:T1} Measurement of the longitudinal relaxation time $T_1$. The data are fitted with a single exponential decaying function, yielding $T_1$ = 5.0 ms.}
\end{figure}

The longitudinal relaxation time $T_1$ of the system was measured by observing the transient decay of the population after initialization.  To eliminate background noise, an alternating measurement was also employed by applying a $\pi$ pulse at 11.36 MHz before the readout. The differential signal obtained from these alternating measurements is shown in Fig.~\ref{fig:T1}. Fitting the data with a single exponential decay yields a longitudinal relaxation time of $T_1$ = 5.0 ms.

\subsection{Dephasing time $T_2^*$}
\begin{figure}[h]
\includegraphics[width=6cm]{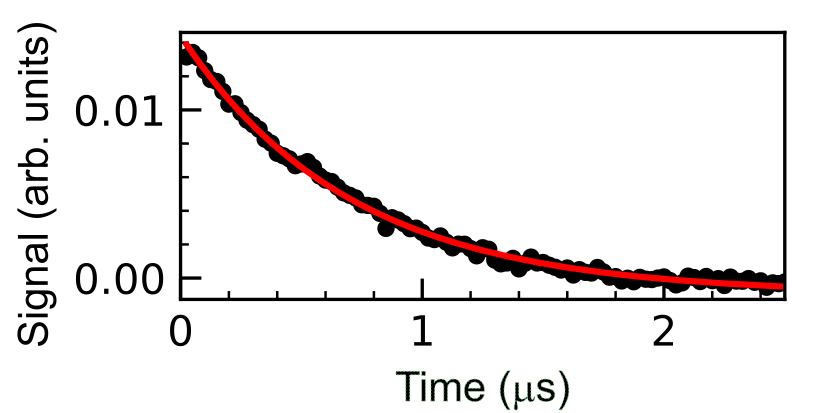}
\caption{\label{fig:Ramsey} Measurement of the dephasing time $T_2^*$ using the Ramsey sequence. The data are fitted with a single exponential decaying function, yielding $T_2^*$ = 701 ns.}
\end{figure}

The dephasing time of the system was measured by observing the transient decay of the system under the Ramsey sequence using soft pulses. After initialization, a weak $\pi/2$ pulse are weakly applied resonantly to the central line to prepare a superposition of the two-level system. The free induction decay was monitored by measuring the coherence through an additional $\pi/2$ pulse, followed by readout of the population. Likewise, alternating measurements were performed by inverting the phase of the final $\pi/2$ pulse. The resulting differential signal is shown in Fig.~\ref{fig:Ramsey}. Fitting with a single exponential decay curve yields a dephasing time of $T_2^*$ = 701 ns. It is worth noting that this measurement was performed under lower laser intensity. A reduction in the dephasing time was observed at higher laser power, which can be attributed to the fluctuations in the zero-field splitting parameter. Therefore, this parameter should be interpreted as the dependence of the sample and the gradient in the bias field.

\subsection{Coherence time $T_2$}
\begin{figure}[h]
\includegraphics[width=6cm]{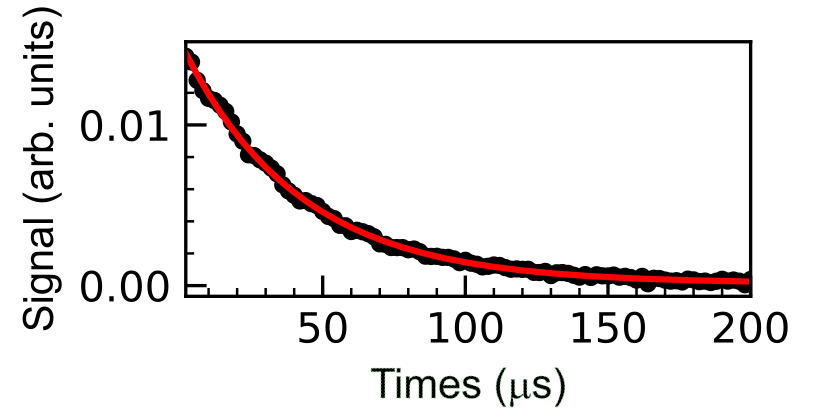}
\caption{\label{fig:echo} Measurement of the Coherence time $T_2$ using the Hahn echo sequence. The data are fitted with a single exponential decay function, yielding $T_2$ = 47 $\mu$s.}
\end{figure}
The coherence time $T_2$ of the system was measured using the Hahn echo sequence. After initialization, a $\pi/2$ pulse was applied resonantly with the central line of the system to create a superposition of the two-level system. An inversion pulse was applied after a delay $\tau$, followed by a second $\pi/2$ pulse after an additional delay $\tau$. Likewise, an alternating measurement was performed by inverting the phase of the final $\pi/2$ pulses. The differential signal is shown in Fig.~\ref{fig:echo}. Fitting the decay with a single exponential function yields a coherence time $T_2$ = 47 $\mu$s.

\section{CCDD with signals}\label{sec:CCDD_sig}
This Appendix details the dynamics of CCDD in the presence of the test target signal in 
Sec.~\ref{sec:magnetometry}
of the main text. First, we present the alternating measurements with $\pm y$ phases for the final $\pi/2$ pulses, which are used to obtain the differential signal in the main text. Then, we summarize the dynamics of the system with different amplitudes for the target signal in CCDD magnetometry. 

\subsection{Transient dynamics}\label{sec:CCDD_sig_alt}
The measurement signals of the transverse CCDD with the test target signal in 
Fig.~\ref{fig_characterization}(c)
of the main text are obtained by subtracting the alternating measurements with $\pm y$ phases for the final $\pi/2$ pulses. 
\begin{figure}[t]
\includegraphics[width=8.6cm]{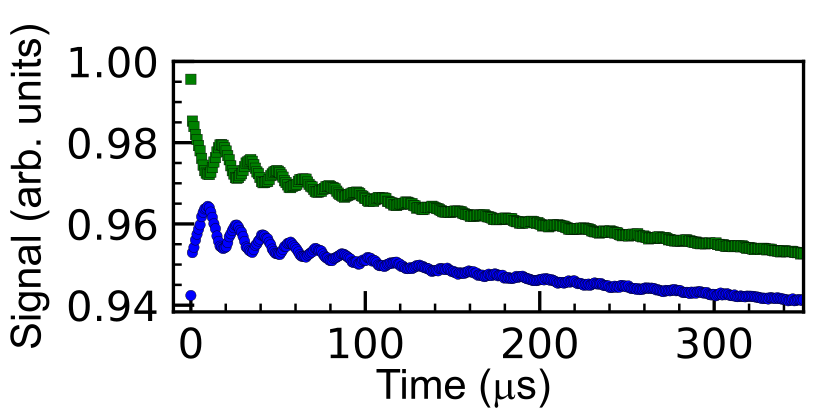}
\caption{\label{fig:alternating} Alternating signals under CCDD with signal (corresponding to 
Fig.~\ref{fig_characterization}(c)
in the main text). Measurement data obtained with $\pi_{+y}/2$ and $\pi_{-y}/2$ before the readout are shown blue circles and green squares, respectively.}
\end{figure}
Figure~\ref{fig:alternating} shows the transient dynamics of the system under CCDD with the target signal for the two alternating measurements. The signals measured with $\pi_{+y}/2$ and $\pi_{-y}/2$ are shown in blue and green, respectively. 

Small oscillations from the MW signal are observed on top of a decaying background. In addition, a gap is present between the alternating measurements. The contrast of the individual oscillations under CCDD with signal is smaller than that of the Rabi oscillations with hard pulse in Fig.~\ref{fig:characterization}(b). We attribute the slowly decaying background decay observed in the individual alternating signals to charge dynamics of the NV center, including conversion between NV$^-$ and NV$^0$. This contribution appears as a common-mode noise and is effectively removed in the differential signal. We performed a series of experiments to study these effects further, which are discussed later.

\begin{figure*}[t]
\centering
\includegraphics[width=17cm]{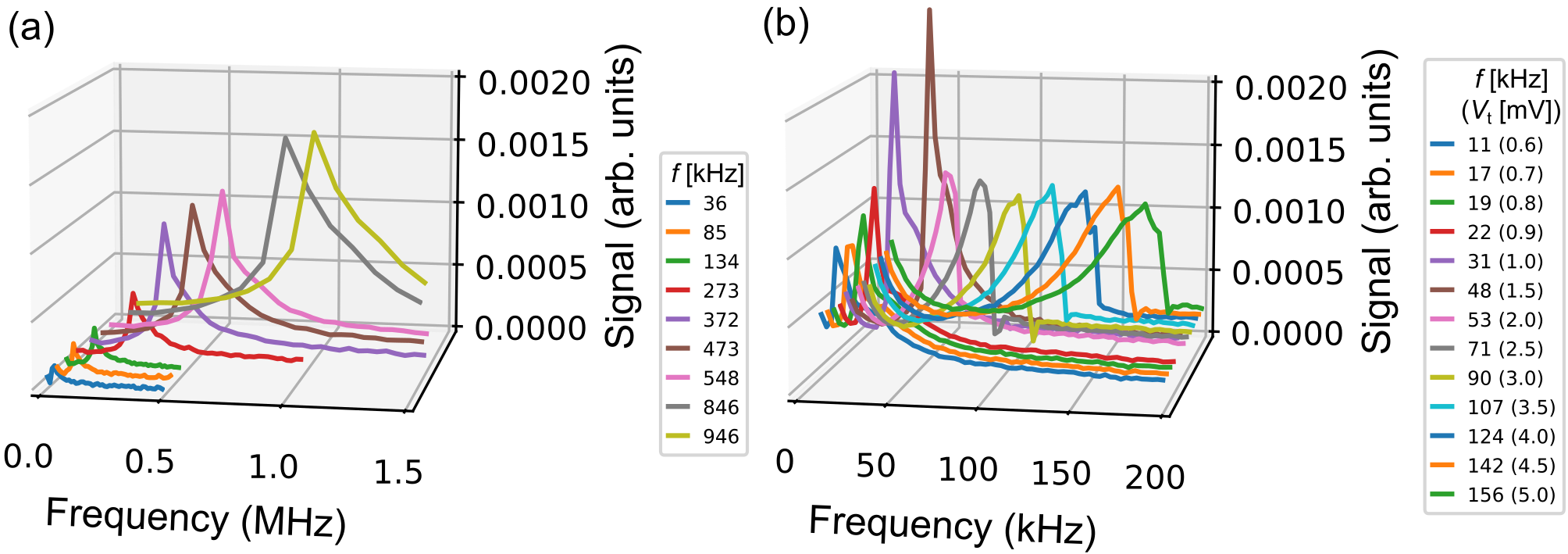}
\caption{\label{fig:CCDD_FT} Fourier transform of the transient measurement at different amplitudes, labelled by the corresponding oscillation frequencies. (a) direct Rabi magnetometry.  
(b) CCDD magnetometry. They are performed by changing the amplitude of input voltage from a MW source $V_{\mathrm{t}}$, denoted in the legend.
}
\end{figure*}

\subsection{Amplitude dependence}

Here, we detail the measuring range comparison between direct Rabi magnetometry and CCDD magnetometry in 
Fig.~\ref{fig_comparison}
of the main text. In the former, Rabi measurements are performed at different amplitudes fed into the PCB board. In the latter, the same experiment as in 
Fig.~\ref{fig_characterization}
is repeated for different source voltage $V_{\rm t}$ of the MW generator. 

To gain insight into the contrast and decay dynamics, the results are shown as Fourier transforms in Fig.~\ref{fig:CCDD_FT}. The Rabi magnetometry spectra in Fig.~\ref{fig:CCDD_FT}(a) exhibit a sharp drop once the Rabi frequency falls below the range of 372-273 kHz, which is close to the linewidth ($\Delta\nu$ = 415 kHz). 
In contrast, the CCDD magnetometry spectra in Fig.~\ref{fig:CCDD_FT}(b) maintain enhanced contrast even below the linedidth. The peak distribution changes drastically with the input voltages. The peak frequency scales linearly with the applied input voltage amplitude, and is later used to calibrate the magnetometry. The width of the distribution increases as the input voltage increases. For some large amplitude above 4 mV, another small peak can be seen. Similar behavior was also observed under the CCDD (see Appendices~\ref{sec:CCDD_rect} and \ref{sec:CCDD_CP}).

For these complex decay dynamics, we do not explore the optimization of the measurement conditions and we demonstrate magnetometry at 2 mV in the main article. Further studies are necessary with various combinations of the first and second drive fields in the CCDD.

\section{Details of magnetometry}\label{sec:magnetometry_details}
This Appendix details the calibration procedure of the amplitude measurement in CCDD magnetometry, corresponding to 
Sec.~\ref{ssec:sensitivity}
of the main text.

\subsection{Conversion of test signal amplitude}
\begin{figure}[h]
\includegraphics[width=8.6cm]{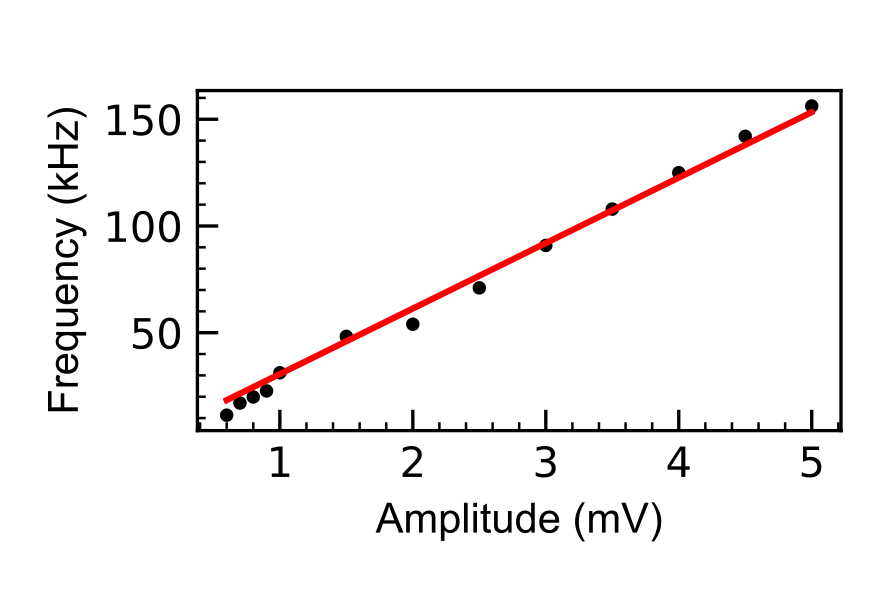}
\caption{\label{fig:freq_amp} Oscillation frequency $\Omega^{\prime}_{\rm t}(=\Omega_{\mathrm{t}}/2)$ under CCDD sequence. The Oscillation frequency $\Omega^{\prime}_{\rm t}$ of the test target microwave signal is measured under the CCDD sequence as a function of its voltage amplitude $V_{\rm t}$ input to the MW generator. A linear fit yields $\Omega_{\rm t}^{\prime}=aV_{\rm t}$ with $a = 30.7 \pm 0.4$ [kHz/mV].
}
\end{figure}

The oscillation frequency $\Omega_{\rm t}^{\prime}$ induced by the target MW signal in CCDD magnetometry is proportional to the signal amplitude. Under the low-attenuation scheme \cite{salhov_protecting_2024}, this relationship is given by $
\Omega_{\rm t}^{\prime} = \frac{1}{2}\Omega_{\rm t} = \frac{1}{2} \gamma_{\rm e} B_{\rm t}
$,     
where $\gamma_{\rm e}$ is the electron gyromagnetic ratio. Figure~\ref {fig:freq_amp} shows the measured oscillation frequencies as a function of the voltage amplitude $V_{\rm t}$ of the test MW signal from the MW source.  A linear fit to the data yields $\Omega_{\rm t}^{\prime} = a V_{\rm t}$, with $a = 30.7 \pm 0.4 $[kHz/mV]. This result gives us 
\begin{equation}
\label{eqB1}
B_{\rm t} = \frac{2a}{\gamma_{\rm e}}V_{\rm t},    
\end{equation}
which is used to convert the input amplitude $V_{\rm t}$ from the MW source to the magnetic field amplitude of the test target signal $B_{\rm t}$.  

\subsection{Response}
\begin{figure}[t]   
\includegraphics[width=8.6cm]{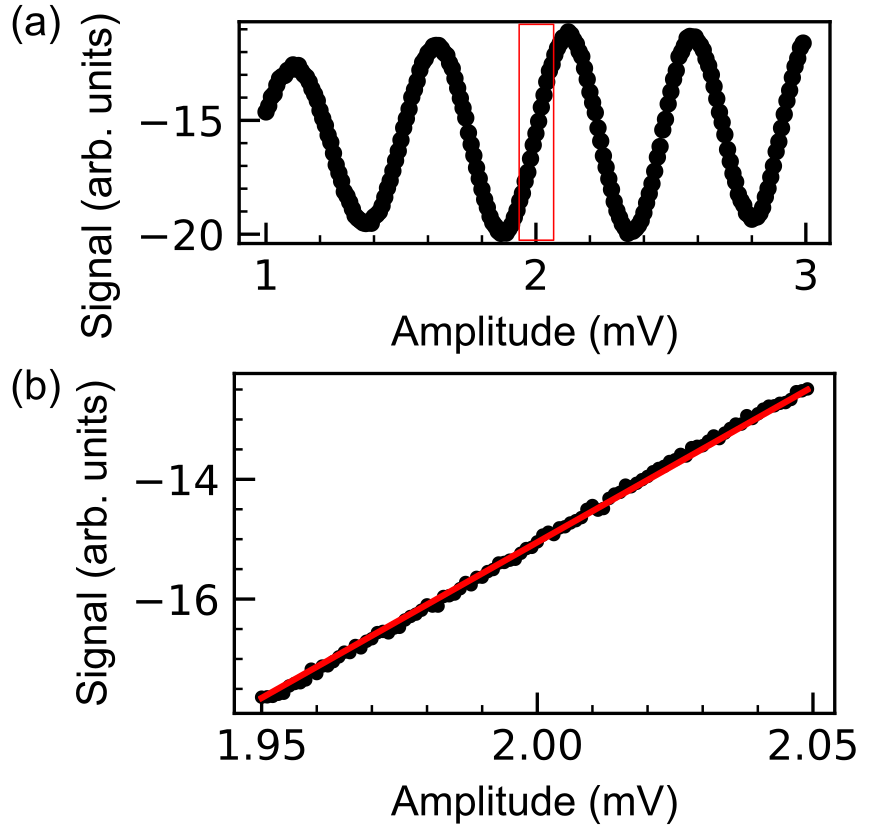}
\caption{\label{fig:response} Calibration of measurement response. (a) Response of the measurement signal $S$ as a function of the voltage amplitude $V_{\rm t}$ of the test target signal, with the interaction time fixed at $\tau$ = 67 $\mu$s. (b) Zoom-in around the linear slope near $V_{\rm t}$ = 2 mV. A linear fit yields $S=RV_{\rm t}$, where $R= 52.0 \pm 0.2$ [/V].}
\end{figure}

To measure the magnetic field amplitude, the measurement signal $S$ must be converted into the corresponding magnetic field amplitude $B$. This is achieved by measuring the response of the system under CCDD as the amplitude of the test signal $V_{\rm t}$ is varied, while keeping the measurement time $\tau$ fixed.  
Figure~\ref{fig:response}(a) shows the response curve when the voltage amplitude $V_{\rm t}$ of the test signal was swept from 1 to 3 mV. The response to the signal $S$ is maximized in the linear region of the sinusoidal curves. This linear regime is shown in Fig.~\ref{fig:response}(b), where the response is described by $S = RV_{\rm t}$ with $R= 52.0 \pm 0.2$ [/V]. In the linear regime, the inverse function can be simply obtained as $V_{\rm t} = S / R$. Combining this with Eq. \ref{eqB1}, we obtain:
\begin{equation}
    B_{\rm t} = \frac{2a}{\gamma_{\rm e}R}S.    
\end{equation}
This relation is used to convert the measurement signal $S$ into the magnetic field amplitude in the sensitivity estimation in the main text. A similar relation is used for the direct Rabi method with $\Omega^{\prime}_{\rm t} = \Omega_1$. It is worth noting that we are interested in the precision of amplitude measurement. A more detailed calibration of the measurement system would be required to improve accuracy.

\subsection{Uncertainty}\label{ssec:uncertainty}
As shown in 
Fig.~\ref{fig_uncertainty}
in the main text, the Allan deviation deviates from $\tau^{-1/2}$ scaling, indicating that our measurements accumulate colored noise. We therefore calculate the measurement uncertainty within the white-noise regime. Since the Allan deviation gradually deviates from $\tau^{-1/2}$ scaling toward a plateau, it is necessary to define the white-noise regime. In the present study, we use the data up to the point where the difference between the Allan deviation and the SEM is below 10 percent of the SEM. We consider that the resulting measured value is reasonably accurate from a metrological point of view. We note that if we use the data up to the plateau, the measured value would be precise, but not necessarily more accurate. A detailed analysis of the measurement accuracy is beyond the scope of the present work.

\section{CCDD with rectangular pulses}\label{sec:CCDD_rect}
In the main text, Rabi frequency of the first drive is set high relative to the linewidth of the system, exceeding $(2\pi)$ 10 MHz. In this section, we examine the behavior of the transverse CCDD using rectangular $\pi/2$ pulses at different drive strengths. These analyses help us to understand the dynamics in Sec.~\ref{sec:CCDD_sig_alt}.

\begin{figure}
\includegraphics[width=8.6cm]{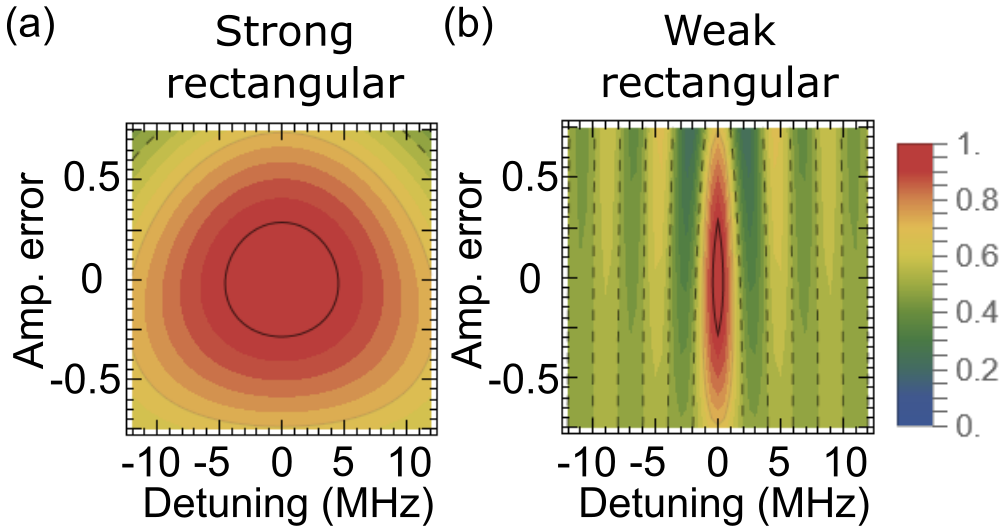}
\caption{\label{fig:fidelity_rect} Fidelity maps of rectangular $\pi/2$ pulses under detuning and amplitude errors. (a) Hard pulse with a Rabi frequency of 10 MHz. (b) Soft pulse with a Rabi frequency of 1 MHz.}
\end{figure}

Here, the same amplitude is used for both $\pi/2$ pulses and the first drive of the CCDD. The behaviors of the CCDD are presented for strong ($\Omega_1 >$  10 MHz  $> A_{\parallel}$) and weak drive fields ($\Omega_1 <  A_{\parallel}$), relative to the hyperfine splittings. Note that in the case of weak drive, $\Omega_1> \Delta \nu$ must still be verified to ensure robust dressing. Strong pulses are expected to robustly excite all the hyperfine lines, while weak pulses primarily address the central line. We simulate the excitation profiles of $\pi/2$ pulses by plotting the fidelity as a function of detuning and amplitude noise, as shown in Fig.~\ref{fig:fidelity_rect}. Experimentally, the drive strength is tuned by adjusting the gain of the microwave amplifier. The ratio between the amplitudes of the first and second drives from the signal generator is kept constant at 10 : 2 such that $\Omega_1 : \Omega_2 \simeq 10 : 1$. 

\begin{figure*}[t]
\includegraphics[width=17cm]{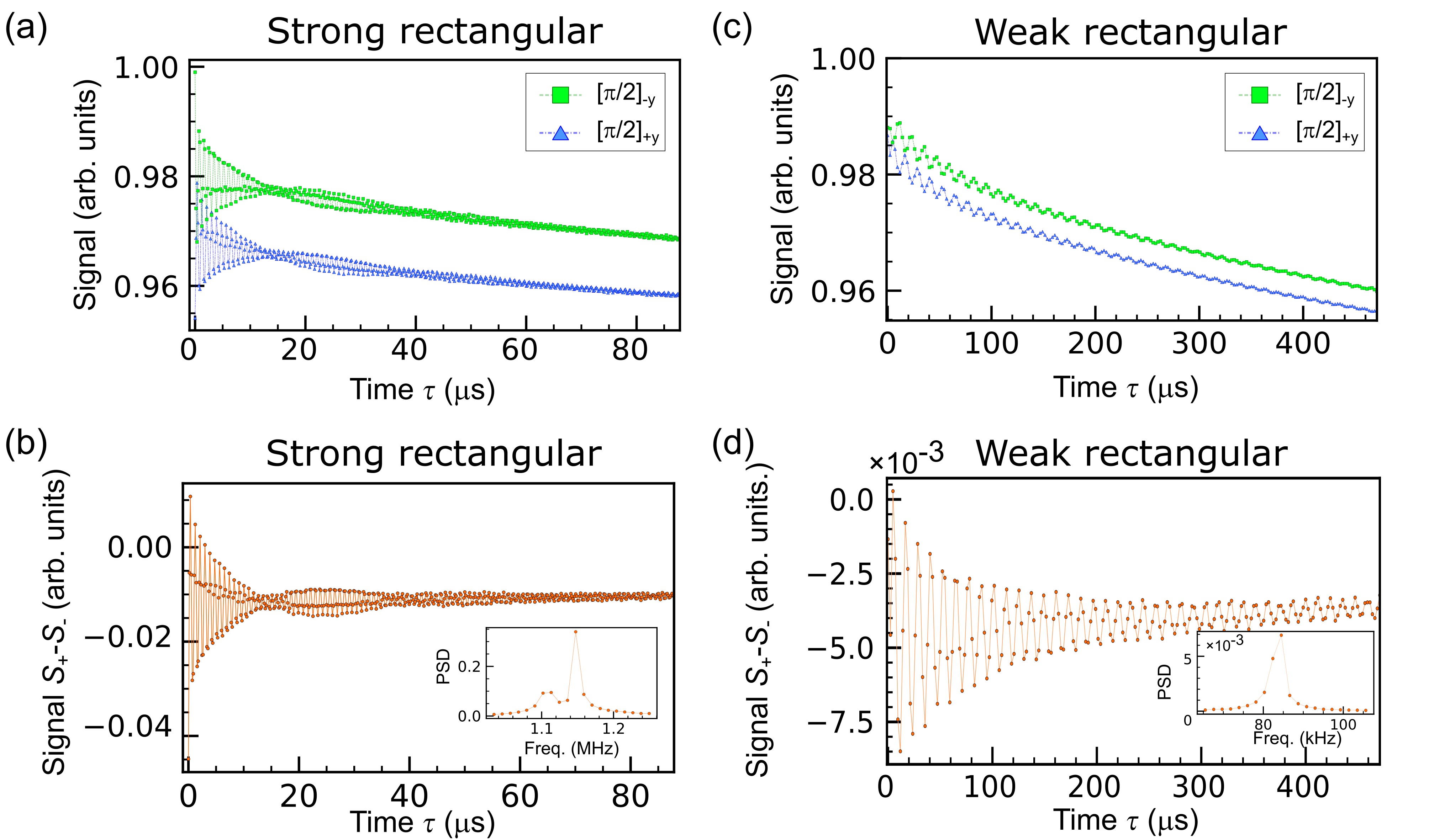}
\caption{\label{fig:CCDD_rect} Transient dynamics of the system under transverse CCDD with rectangular $\pi/2$ pulses at different drive strengths. (a), (b) Alternating measurements and corresponding differential signals for the strong drive case ($\Omega_1$ = $(2\pi)$11.36 MHz). The power spectrum density (PSD) of the differential signal is represented in the inset of (b). The differential signal is fitted with $D=D_{\rm osc} + D_{\rm b}$, where the oscillatory component is  given by 
$D_{\rm osc}=\sum_{n=1,2}\{A_{\Omega_2}^{(n)} \exp[-t/T_{\Omega_2}^{(n)}] \cos(\Omega_2^{(n)}t+\phi_{\Omega_2}^{(n)}\}$
and the background is described by
$D_{b}=B_{\Omega_2}\exp[-t/T_{{\rm b}, \Omega_2}] + C_{\Omega_2}$. Detailed fitting parameters are provided in 
Appendix~\ref{sec:fit}.
(c), (d) Corresponding measurements and the differential signal for the weak drive case ($\Omega_1$ = $(2\pi)$846 kHz). The amplitude of the control fields is adjusted via the gain of a MW amplifier, maintaining a constant ratio between the first and second drives. }
\end{figure*}

\subsection{Strong drive}
Fig.~\ref{fig:CCDD_rect}(a) shows the CCDD signal obtained using a strong drive $\Omega_1$ = $(2\pi)$11.36 MHz. Green circles and blue triangles represent measurements taken with $-y$ and $+y$ phase for the  $\pi/2$ pulses, respectively. Note that the time step is set to integer multiples of the Rabi period $\tau_{\Omega_1} = 2\pi / \Omega_1$ to move into the rotating frame. Slow oscillations are observed on top of a common-mode decaying background. The subtraction of these alternating-phase signals can be used to remove this background noise, which is shown in 
Fig.~\ref{fig:CCDD_rect}(b)
with its power spectrum density (PSD) in the inset. As expected, slower oscillations are observed at a frequency of 1.15 MHz, which agrees with $\Omega_1/10$, corresponding to the second drive. The decay time is prolonged to $\mu$s order compared to that of the single drive $T_{\Omega_1}$, due to the suppression of the amplitude noise. 

However, several features emerge that have received limited attention in previous studies on the CCDD using single NV centers or a homogeneous ensemble of NV centers. 
These include
\begin{enumerate*}[label=(\roman*), itemjoin={{, }}, itemjoin*={{, and }}]
    \item \label{feat:gap} a gap between the alternating-phase signals
    \item \label{feat:amp} a reduction in the measurement-signal amplitude
    \item \label{feat:bg} a saturating background in addition to the oscillatory component
    \item \label{feat:osc} additional weak oscillations at a frequency of approximately 1.11 MHz
\end{enumerate*}.

To quantitatively capture these anomalies, we model the differential signal as a sum of damped oscillatory terms and a saturating background:
\begin{align}    
\label{eq:D_CCDD} 
D&=D_{\rm osc} + D_{\rm b} \\
D_{\rm osc}&=\sum_{n=1,2}
\left[A_{\Omega_2}^{(n)} 
\exp \left[-\left(t / T_{\Omega_2}^{(n)} \right) ^{p_{\Omega_2}^{(n)}} \right] \cos\left( \Omega_2^{(n)}t+\phi_{\Omega_2}^{(n)} \right)\right]
\nonumber
\\
D_{\rm b} &= B_{\Omega_2}\exp(-t/\tau_{\Omega_2}) + C_{\Omega_2}. \nonumber
\end{align}
The oscillatory part $D_{\rm osc}$ is well described by a combination of primary oscillations with a larger amplitude $A_{\Omega_2}^{(1)}$ = 0.0249 at $\Omega_2^{(1)}$ = $(2\pi)$1.143 MHz, and secondary oscillations with a smaller amplitude $A_2^{(1)}$ = 0.014 at $\Omega_2^{(1)}$ = $(2\pi)$1.130 MHz. The former follows a single exponential decay ($p_{\Omega_2}^{(1)}$ = 1), while the latter is fitted well by a Gaussian decay ($p_{\Omega_2}^{(2)}$ = 2). 
The amplitudes of the primary and secondary oscillations in the differential mode component $D/2$ are $A_{\Omega_1}/2$ = 0.0125 and $A_{\Omega_2}/2$ =  0.0070, respectively, which is lower than that of the first drive $A_{\Omega_1}$ = 0.0287 [feature~\ref{feat:amp}]. This suggests that not all oscillations generated by the first drive contribute to the second oscillations. The frequency of the secondary peak $\Omega_2^{(2)}$=$(2\pi)$ 1.130 MHz [feature~\ref{feat:osc}] does not agree with that expected from Mollow side bands \cite{mollow_power_1969, joas_quantum_2017, wang_coherence_2020}. These results raise concerns when applied to GHz-range magnetometry, where the target MW signal must be resonant with the energy levels of the dressed states. Furthermore, it remains unclear whether this secondary peak can be driven by the target microwave field. 

The background $D_{\rm b}$ describes the saturating signal [feature~\ref{feat:bg}], which asymptotically approaches $C_{\Omega_2}$. This is more pronounced when the measurement time is extended (see Sec.~\ref{sec:bg}). The observed gap between the alternating signals [feature~\ref{feat:gap}] manifests as the deviation of the differential background constant from 0: $C_{\Omega_2} = -0.012$. In standard pulsed measurements such as Ramsey or Hahn echo, the alternating signals typically converge after the relaxation time. For sensing purposes, the presence of this gap necessitates a larger dynamic range in the analog-digital converter (ADC) to accurately measure the alternating signals. This approach may not be preferable when the detection range of the ADC must be reduced to enhance the bit resolution of the measurements \cite{barry_sensitive_2024}. While it remains unclear from the current data whether the saturating background has settled, we now turn to compare these results with those obtained under a weak drive to further investigate the origin of these features. A detailed discussion of the background behavior will follow later.

\subsection{Weak drive}

Let us now examine the transverse CCDD with a weak $\pi/2$ drive. The strength of the drive fields is reduced by decreasing the gain of the MW amplifier. 
\begin{figure}[t]
\includegraphics{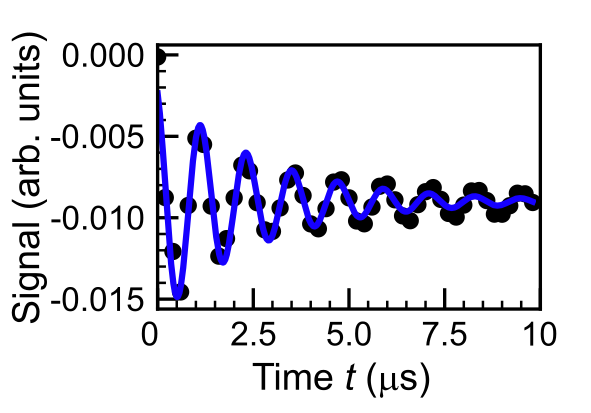}
\caption{\label{fig:Rabi_weak}Rabi oscillations under weak driving. The MW drive field is applied on resonance with the central frequency $\omega_0$ of the sublevels. The signal is fitted with $A_{\Omega_1}\exp[-(\tau/T_{\Omega_1})]\cos[\Omega_1 t] + C_{\Omega_1}$, yielding $\Omega_1$ = $(2\pi)$837 kHz, $T_{\Omega_1}$ = 2.67 $\mu$s and $A_{\Omega_1} = 0.0071$.}
\end{figure}
The result of the Rabi oscillation under weak drive is shown in Fig.~\ref{fig:Rabi_weak}. The signal was fitted with an exponentially decaying cosine function, yielding a Rabi frequency of $\Omega_1$ = $(2\pi)$ 837 kHz and a decay time of $T_{\Omega_1}$ = 2.67 $\mu$s. The reduced amplitude $A_{\Omega_1} = 0.0071$, compared to that under strong driving, suggests that the excitation predominantly targets the  central line of the hyperfine sublevels.

Figure~\ref{fig:CCDD_rect}(c) shows the alternating signals from the CCDD with reduced drive amplitude. As in the strong-drive case, a gap remains between the two signals. The corresponding differential signal in Fig.~\ref{fig:CCDD_rect}(d) exhibits oscillations at approximately 85 kHz, consistent with $\Omega_1/10$. Compared to the strong-drive case, the initial decay is more gradual, 
and the additional peak [feature~\ref{feat:osc}] is not present beyond the one associated with these oscillations. An exponentially saturating background is clearly present. Measurements over extended timescale reveals that this background approaches zero as the alternating signals merge over time (see Sec.~\ref{sec:bg}). Therefore, the gap in the alternating signals manifest as the decaying background in the differential signal.

Similarly, this differential signal is again well described by the sum of damped oscillations and a saturating background in Eq. \ref{eq:D_CCDD}. Remarkably, the oscillation amplitude $A_{\Omega_2}$ is as low as 3.3 $\times10^{-3}$, attributed to the reduced contrast of the first drive $A_{\Omega_1}$ = 7.1 $\times10^{-3}$. The gap between the alternating signals is still evident with $C_{\Omega_2}$ = -4.6 $\times10^{-3}$. The characteristic time of the saturating background $\tau_{\Omega_2}$ is 471 $\mu$s.

The absence of the secondary peak under the weak drive suggests that the signal observed with the strong drives consists of a combination of primary oscillations from the central line and weaker ones mainly from the sidebands. Under the strong drive, all three lines are excited, but the detuned sidebands are weakly excited and oscillate at different frequencies. In contrast, under the weak drive, the central line is mainly excited, and the contributions from the sidebands are minimal, resulting in single-tone oscillations. 

Even considering the reduced contributions from the sidebands under the weak drive, the contrast of the subtracted signal $A_{\Omega_2} = 3.3 \times10^{-3}$ remains smaller than that of the primary oscillations $A_{\Omega_2} = 0.024$, which is presumed to originate mainly from the central line. We attribute this discrepancy to a temporal shift of the energy level. In our system, the energy gap is given by $\omega_0 = D - \gamma B$, where $D$ is the zero-field splitting of spin sublevels of the NV center, $\gamma$ is the electron gyromagnetic ratio and $B$ is the bias static magnetic field applied along the N-V axis. The zero-field splitting varies linearly with the temperature as $D(T) = D_0 + \frac{dD}{dT}T$, with $D_0$ = $(2\pi)$2.87 GHz and $\frac{dD}{dT}$ = - 74.2 kHz/K \cite{acosta_temperature_2010, doherty_temperature_2014}. Laser-induced heating inside the diamond is likely to make the resonant frequency fluctuate in the course of measurements. Therefore, we consider that it is preferable to work under strong drive conditions to beat this fluctuation. Having analyzed the oscillatory parts $D_{\rm osc}$ in detail, we now investigate the background component $D_{\rm b}$. 

\subsection{Background}\label{sec:bg}

\begin{figure}[t]
\includegraphics[width=8.6cm]{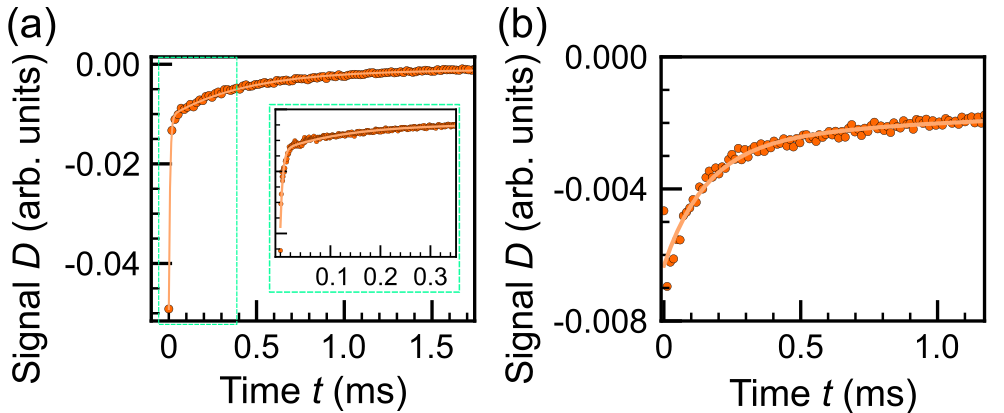}
\caption{\label{fig:CCDD_background}Transverse CCDD with (a) strong and (b) weak drive over extended timescales. The inset of (a) zooms up its initial region, indicated by the light-green dashed square. Each signal exhibits an initial rapid decay followed by a slower decay.  Both of them are well described by a bi-exponential saturation function: $D = \sum_{n=1,2}b_{\Omega_2}^{(n)} \exp[-t/t_{\Omega_2}^{(n)}] + c_{\Omega_2}
$. For the strong drive, the initial and subsequent decay times are $t_{{\Omega_2}, {\rm s}}^{(1)}$ = 4.75 $\mu$s and $t_{{\Omega_2}, {\rm s}}^{(2)}$ = 444 $\mu$s, respectively; for the weak drive, $t_{{\Omega_2}, {\rm w}}^{(1)}$ = 153 $\mu$s and $t_{{\Omega_2}, {\rm w}}^{(2)}$ = 1.03 ms}
\end{figure}

Figure~\ref{fig:CCDD_background}(a) and (b) show the differential signals of the transverse CCDD under strong and weak driving, respectively, measured over extended timescales. Note that the time step is set as integer multiples of the CCDD period $T_{\Omega_2}$. Both results exhibit an initial rapid decay followed by a slower decay. The former should contain the envelope of the oscillatory component $D_{\rm osc}$, while the latter should correspond to the background decay $D_{\rm b}$. The background signals saturates at long times. This behavior can be well captured by a bi-exponential fit:

\begin{equation}\label{eq:D_spin_lock}
D = \sum_{n=1,2}b_{\Omega_2}^{(n)} \exp[-t/t_{\Omega_2}^{(n)}] + c_{\Omega_2}
\end{equation}

The characteristic time of the initial decay $t_{ \Omega_2}^{(1)}$ for the CCDD with strong and weak drives are 12.0 $\mu$s and 149 $\mu$s, respectively. 
As expected, these timescales are consistent with the decay time of the second drive $T_{\Omega_2}$. 
In contrast, the characteristic times for the subsequent decay $t_{\Omega_2}^{(2)}$ is 444 $\mu$s and 1.03 ms for strong and weak drives, respectively. Both values are shorter than the longitudinal relaxation time of the system $T_1$ = 5 ms. 

In summary, the background signal [feature~\ref{feat:bg}] in each alternating signal decays at timescales longer than the oscillation period of the CCDD, which manifests as a gap between the alternating signals [feature~\ref{feat:gap}] at the timescale of CCDD oscillations.

\section{CCDD with composite pulses}\label{sec:CCDD_CP}

\begin{figure}[t]
\includegraphics[width=8.6cm]{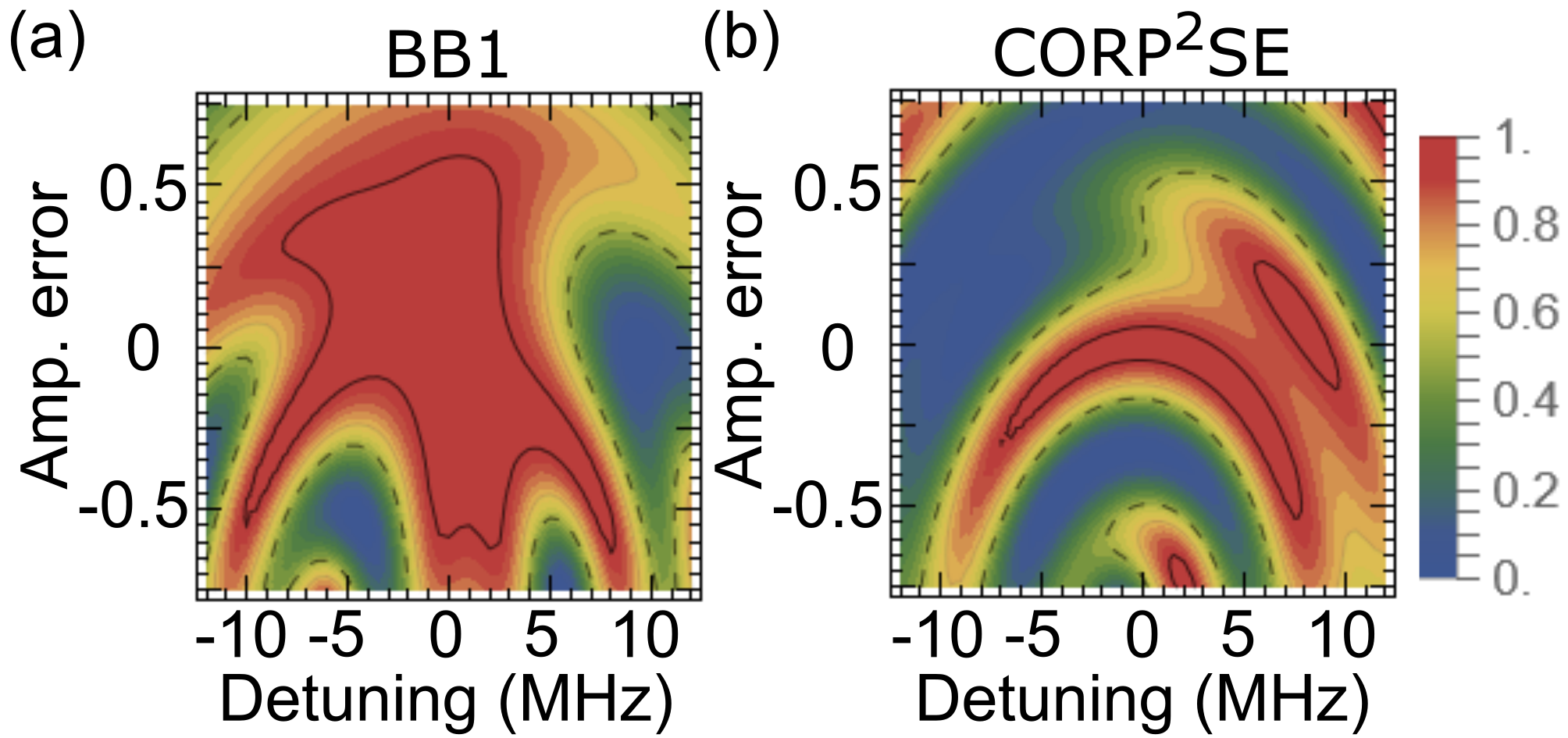}
\caption{\label{fig:fidelity_CP} Fidelity plots of composite $\pi/2$ pulses, consisting of rectangular pulses with Rabi frequency of $\Omega_1=(2\pi)\, 10$ MHz under detuning and amplitude errors. (a) BB1 composite pulses, demonstrating robustness agains amplitude noise (plse length error). (b) CORP$^2$SE composite pulse, showing robustness primarily against detuning (off-resonance error), with reduced fidelity under large amplitude errors.}
\end{figure}

The observations  
presented  in Appendix~\ref{sec:CCDD_rect} for transverse CCDD with rectangular $\pi/2$ pulses
indicate a possible link of the transient dynamics of the transverse CCDD to the excitation characteristics of the $\pi$/2 pulses. To test this hypothesis, we replace the rectangular pulses in Fig.~1(b) with a set of composite pulses (CPs) \cite{levitt_composite_1986, vandersypen_nmr_2005,merrill_progress_2014, suter_colloquium_2016} that create several initial states with different deviations dependent on detuning and ampllitude noise. We then examine how these varied deviations affect the subsequent CCDD dynamics. In this section, we focus on two composite $\pi/2$ pulses: BB1 \cite{wimperis_broadband_1994} and CORP$^2$SE \cite{kukita_geometric_2022}. 

BB1 is well known as a broadband composite pulse that is robust against pulse length error (PLE) \cite{wimperis_broadband_1994, merrill_progress_2014}. A BB1 CP for a rotation angle $\theta$ with phase $\phi$ is composed of four pulses:
\begin{equation}
    {\rm BB1}(\theta, \phi): (\pi)_{\phi+\phi_1} (2\pi)_{\phi+\phi_2} (\pi)_{\phi+\phi_1} (\theta)_{\phi},
\end{equation}
where $\phi_1 = \cos^{-1}(-\theta/4\pi)$ and $\phi_2 = 3\phi_1$.
The fidelity plot of a BB1 $\pi/2$ at a Rabi frequency of 10 MHz in Fig.~\ref{fig:fidelity_CP}(a) exhibits high uniformity along the amplitude noise axis, confirming its robustness against PLE. It also shows partial robustness against off-resonance error (ORE). Since the excitation profile of BB1 is not drastically different from that of rectangular pulses, it serves as an example of a more robust pulse. 

On the other hand, CORP$^2$SE \cite{kukita_geometric_2022} is a recently developed CP based on the CORPSE CP, which is known for its robustness against ORE \cite{cummins_tackling_2003, merrill_progress_2014, kukita_geometric_2022}. A CORP$^2$SE CP consists of three pulses:
\begin{equation}
{\rm CORP^2SE}(\theta, \phi): (\theta_1)_{\phi-\frac{3\pi}{4}}(\theta_2)_{\phi-\frac{\pi}{4}}(\theta_1)_{\phi-\frac{3\pi}{4}}.
\end{equation}
Here, $\theta_1$ and $\theta_2$ are given by
\begin{eqnarray}
    \theta_1 = \sin^{-1} \left(-\frac{1-\alpha^2}{1+\alpha^2} \right), 
    \theta_2 = \cos^{-1}(\alpha^2)
\end{eqnarray}
where $\alpha = \cos(\theta / 2)$. Figure~\ref{fig:fidelity_CP}(b) shows the fidelity plot of CORP$^2$SE $\pi/2$. It exhibits uniformity more along the detuning axis, confirming its robustness against ORE. However, its fidelity drops rapidly with increasing amplitude noise. This is helpful to focus on the effects of broadband excitations, independent of robustness against amplitude noise. 

\subsection{BB1}
\begin{figure*}[]
\includegraphics[width=17cm]{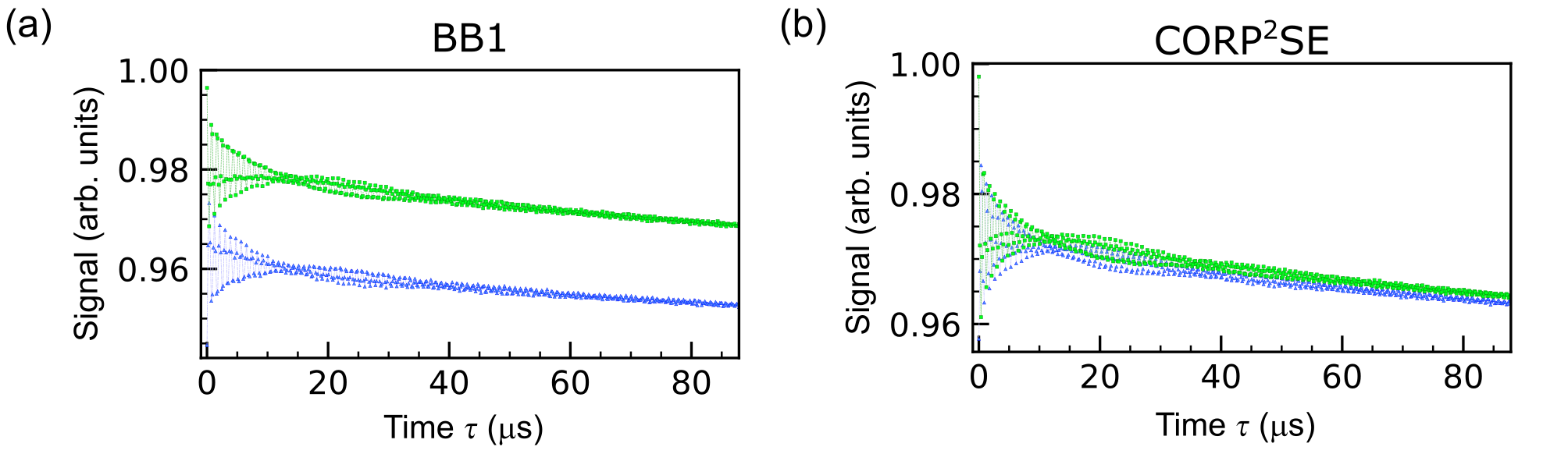}
\caption{\label{fig:CCDD_CP} Transient dynamics of the system under transverse CCDD with composite pulses (CPs): (a) BB1 and (b) CORP$^2$SE. Each composite pulse is implemented by changing the pulse duration and phases while keeping the Rabi frequency at $\Omega_1$ = $(2\pi)$11.36 MHz.}
\end{figure*}

The broadband excitation of BB1 is expected to excite all hyperfine lines with high fidelity. Moreover, its robustness against PLE should precisely prepare the superposition state even under inhomogeneous drive fields, leading to firm locking to the drive field. The results of CCDD with BB1 CPs are displayed in Fig.~\ref{fig:CCDD_CP}(a). Notably, the gap between the alternating signals persists. This result suggests that the gap is unlikely to be associated with the fidelity of the $\pi/2$ pulse. No significant differences are observed in the differential signal (see Fig.~\ref{fig:CCDD_fit}). 
Therefore, both hard rectangular pulses and BB1 CPs can excite the system in a similar way, and the corresponding time dynamics under the CCDD are similar. In short, the improved fidelity does not eliminate the observed gap.

\subsection{CORP$^2$SE}
Let us now turn to CORP$^2$SE. While CORP$^2$SE CPs are also expected to excite also all hyperfine lines, they are more susceptible to amplitude noise originating from the inhomogeneous drive fields. Figure~\ref{fig:CCDD_CP}(b) shows the transverse CCDD combined with CORP$^2$SE CP. Intriguingly, the gap between the alternating signals [feature~\ref{feat:gap}] is significantly reduced. The alternating signals become more symmetric with respect to a monotonically decreasing background signal. 
No significant differences are observed in the differential signal except for a reduction in the background offset to $C_{\Omega_2}$=-0.0016, which corresponds to the reduced gap according to the previous discussion in Sec.~\ref{sec:bg}.

 Given the moderate robustness of CORP$^2$SE, this result supports the notion that the gap is not related to the fidelity. 
 The behavior of the CCDD sequence depends on the prepared initial state. In transverse CCDD, the system is initialized in an $x$-state, orthogonal to the direction of the first drive, satisfying the spin-locking condition. 
 Our fidelity plots characterize the precision of state preparation, but do not indicate how the states deviate in the case of low fidelity.
 The main behavior of the CCDD can be largely attributed to the central line, which is prepared near the $x$-state and contributes to oscillations driven by the second drive. In contrast, the hyperfine sidebands exhibit markedly different behavior depending on their prepared states. All $\pi/2$ pulses with strong drives robustly excite the central line, but only the CORP$^2$SE may cancel the unwanted transformation of the sideband states, mitigating the contribution to the gap. 
We investigate this hypothesis by removing the second drive from the transverse CCDD, which corresponds the spin-locking experiment 
(see Appendix~\ref{sec:spin_lock}).

\subsection{CORPSE}\label{sec:CCDD_CORPSE}
\begin{figure}[h]
\includegraphics[width=8.6cm]{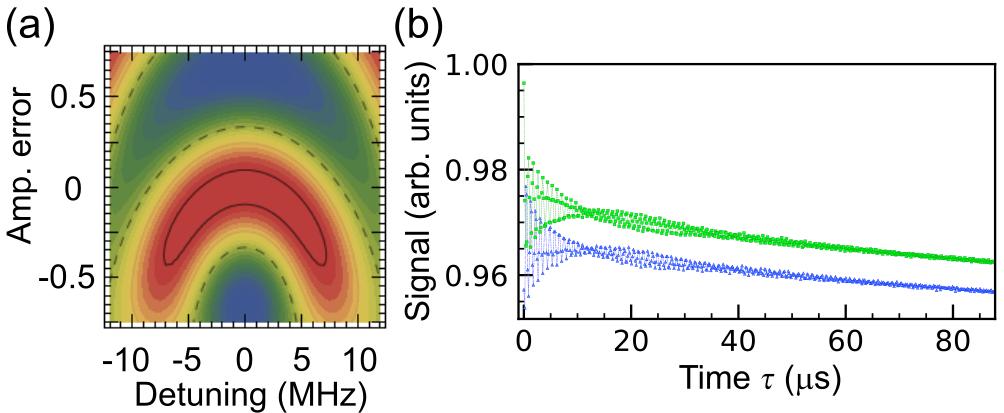}
\caption{\label{fig:CORPSE} (a) Fidelity map of the CORPSE CP under detuning and amplitude errors, showing robustness primarily against detuning (off-resonance errors). (b) Transient dynamics of the system under transverse CCDD with CORPSE CP. Compared to CORP$^2$SE, only a minor reduction in the signal gap is observed, suggesting that the excitation profile of the CORPSE family is not the main reason for the reduced gap.}
\end{figure}

CORP$^2$SE is a family of the CORPSE pulse family, known for its robustness against off-resonance errors \cite{cummins_tackling_2003}. The CORPSE pulse consists of three pulses:
\begin{equation}
{\rm CORPSE}(\theta, \phi): (\theta_1)_{\phi}(\theta_2)_{\phi-\pi}(\theta_3)_{\phi},
\end{equation}
where the rotations angles are given using the parameter $\beta = \sin^{-1}[\frac{1}{2}\sin(\frac{\theta}{2})]$ as follows:
\begin{align}
\theta_1 & = 2n_1\pi + \frac{\theta}{2}-\beta \\
\theta_2 & = 2n_2\pi-2\beta \\
\theta_3 & = 2n_3\pi + \frac{\theta}{2}-\beta 
\end{align}
with $n_1$, $n_2$, and $n_3$ being integers. The robustness of CORPSE against ORE can be seen from the fidelity map plotted against detuning and amplitude error in Fig.~\ref{fig:CORPSE}(a). The corresponding result of the CCDD is shown in Fig.~\ref{fig:CORPSE}(b). Although the gap is slightly reduced, the effect is not significant in comparison with the CORP$^2$SE.

\section{Spin-locking}\label{sec:spin_lock}

\begin{figure}[t]
\includegraphics[width=6cm]{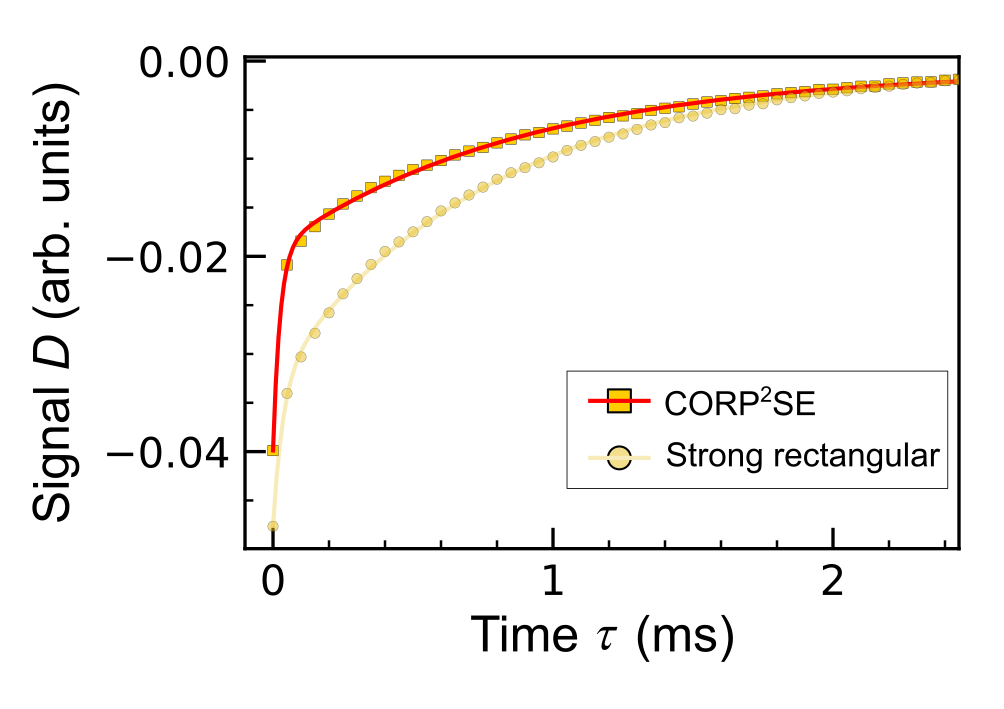}
\caption{\label{fig:spin_lock_comparison}Spin locking with CORP$^2$SE CP (squares), compared with the result using the strong rectangular pulses (circles). The signal under CORP$^2$SE decays more rapidly. 
The signals are fitted with a bi-exponential saturation curve, yielding decay time of 
$t_{\rm sl, s}^{(1)}$ = 36.4 $\mu$s and $t_{\rm sl, s}^{(2)}$ = 786 $\mu$s for the strong rectangular pulses, and $T_{\rm sl, C2}^{(1)}$ = 25.1 $\mu$s and $T_{\rm sl, C2}^{(2)}$ = 888 $\mu$s for the CORP$^2$SE. 
}
\end{figure}
The spin-locking sequence is performed similarly to the transverse CCDD in 
Fig.~\ref{fig_diagram}(b)
of the main text, except that the second drive is not applied. 

\subsection{CORP$^2$SE and Rectangular pulse}\label{sec:spin_lock_c2_rect}
As discussed in the previous section, the CORP$^2$SE CP has a distinctive effect on the background in transverse CCDD. To investigate this effect further, we now apply CORP$^2$SE to a spin-locking sequence and compare the resulting dynamics with those obtained using strong rectangular pulses. 
The differential signals with CORP$^2$SE $\pi$/2 pulses and strong rectangular pulses are presented in Fig.~\ref{fig:spin_lock_comparison}. Both traces exhibit a dual-time decay and gradually approach zero. Interestingly, the signal with CORP$^2$SE decays more rapidly at short times. A bi-exponential saturation curve yields characteristic decay times of $T_{\rm sl, C2}^{(1)}$ = 25.1 $\mu$s and $T_{\rm sl, C2}^{(2)}$ = 888 $\mu$s for the CORP$^2$SE, and $t_{\rm sl, s}^{(1)}$ = 36.4 $\mu$s and $t_{\rm sl, s}^{(2)}$ = 786 $\mu$s for the strong rectangular pulses.

This result suggests that two types of subsystems follow different types of decay dynamics that contribute to the initial decay and subsequent decay. The CORP$^2$SE appears to enhance the initial decay. We attribute the subsequent long-lived component to the on-resonant central lines, while the fast initial decay likely arises from the hyperfine-detuned sidebands. We consider that different initial state distributions determined by the excitation profile of the $\pi/2$ pulses create the difference in this effect. Notably, a similar dual-time decay was also observed in spin-locking experiments under strong drive \cite{nomura_near-field_2021}.

\subsection{Strong and weak drives}

\begin{figure}[t]
\centering
\includegraphics[width=8.5cm]{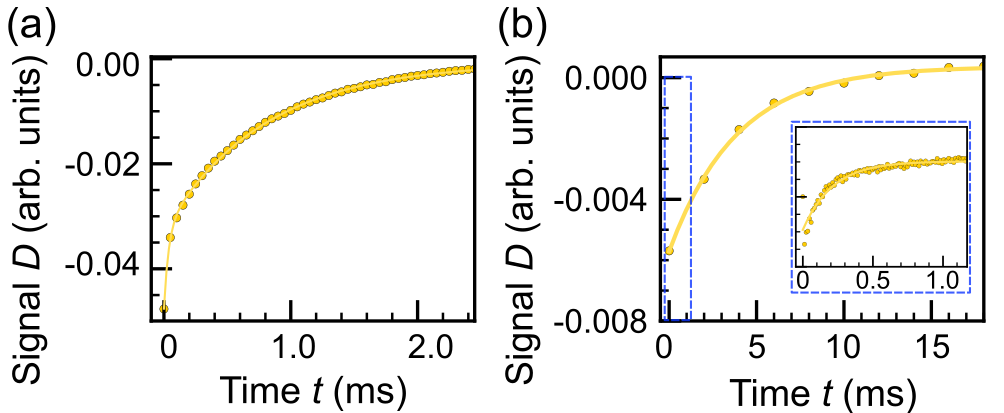}
\caption
{\label{fig:spin_lock}Spin locking with (a) strong and (b) weak drive. The inset in (b) zooms up the initial part, indicated by the blue dashed square. Both datasets are characterized using bi-exponential saturation curves, following the same fitting model in Fig.~\ref{fig:CCDD_background}. For the strong drive, the signal is fitted with a bi-exponential saturation curve, yielding decay time of 
$t_{\rm sl, s}^{(1)}$ = 36.4 $\mu$s and $t_{\rm sl, s}^{(2)}$ = 786 $\mu$s. For the weak drive, the signal is fitted with a single-exponential saturation function in two different timescales, yielding 
$t_{\rm sl, w}^{(1)}$ = 212 $\mu$s and $t_{\rm sl, w}^{(2)}$ = 3.87 ms.}
\end{figure}

We next investigate the spin-locking dynamics under strong and weak drives to probe the decay dynamics in the transverse CCDD sequence in Fig.~\ref{fig:CCDD_background}. 
Figure~\ref{fig:spin_lock} (a) and (b) display the results of the spin-locking sequence under strong and weak drives, respectively. Both traces exhibit dual-time decay and asymptotically approach 0. The signal from the strong drive in Fig.~\ref{fig:spin_lock} is well described by a bi-exponential function, yielding $t_{\rm sl, s}^{(1)}$ = 36.4 $\mu$s and $t_{\rm sl, s}^{(2)}$ = 786 $\mu$s. For the weak drive, single-exponential fits applied to the short- and long- timescale regions give $t_{\rm sl, w}^{(1)}$ = 212 $\mu$s (inset) and $t_{\rm sl, w}^{(2)}$ = 3.87 ms, respectively.

These results provide insight into the background signal in transverse CCDD in Fig.~\ref{fig:CCDD_background}. Remarkably, the transient dynamics of the spin locking under weak drive [inset of Fig.~\ref{fig:spin_lock}(b)] resemble the decay behavior of transverse CCDD under weak drive [in Fig.~\ref{fig:CCDD_background}(b)], while the dynamics under strong drive differ substantially. The corresponding time scales extracted from CCDD and spin-locking experiments for strong and weak drives are summarized 
in Tables~\ref{tab:summary_time_strong} and \ref{tab:summary_time_weak}, respectively. 

\begin{table}[h]
    \centering
    \begin{tabular}{cccc}
    \hline
    \hline
    
    & $T_{\Omega_2}^{(1)}$  
    & $T_{\Omega_2}^{(1)}$ 
    & 
    \\
    \hline
    \makecell[c]{CCDD \\ (fine timescale)} & 12.0 $\mu$s & 15.5 $\mu$s  & 
    \\
    \hline
    &&&
    \\
    \hline
    \hline
    
    &\multicolumn{2}{c}{$t_{\Omega_2}^{(1)}$}   
    &$t_{\Omega_2}^{(2)}$  
    \\
    \hline
    \makecell[c]{CCDD \\ (coarse timescale)}
     & \multicolumn{2}{c}{4.75 $\mu$s} & 444 $\mu$s
     \\
    \hline
    &&&
    \\
    \hline
    \hline
    
    &\multicolumn{2}{c}{$t_{\rm sl, s}^{(1)}$}   
    &$t_{\rm sl, s}^{(2)}$  
    \\
    \hline
    Spin lock & \multicolumn{2}{c}{36.4 $\mu$s} & 786 $\mu$s
    \\
    \hline
    \end{tabular}
\caption[Summary of corresponding timescales with a strong drive]{Summary of corresponding timescales with a strong drive. Top: decay times of initial oscillations of the CCDD at fine timescales in Fig.~\ref{fig:CCDD_rect}
Middle: decay time of the saturating curve of the CCDD at coarse timescales in Fig.~\ref{fig:CCDD_background}. 
Bottom: decay time of the spin locking in Fig.~\ref{fig:spin_lock}.
}
    \label{tab:summary_time_strong}
\end{table}

\begin{table}[h]
    \centering
    \begin{tabular}{ccc}
    \hline
    \hline
    
    & $T_{\Omega_2}$  
    & 
    \\
    \hline
    \makecell[c]{CCDD \\ (fine timescale)} & 149 $\mu$s  & 
    \\
    \hline
    &&
    \\
    \hline
    \hline
    
    &$t_{\Omega_2}^{(1)}$   
    &$t_{\Omega_2}^{(2)}$  
    \\
    \hline
    \makecell[c]{CCDD \\ (coarse timescale)}
     & 153 $\mu$s & 1.03 ms
     \\
    \hline
    &&
    \\
    \hline
    \hline
    
    &$t_{\rm sl, w}^{(1)}$   
    &$t_{\rm sl, w}^{(2)}$  
    \\
    \hline
    Spin lock & 212 $\mu$s & 3.87 ms
    \\
    \hline
    \end{tabular}
    \caption[Summary of corresponding timescales with a weak drive]{Summary of corresponding timescales with a weak drive. Top: decay times of initial oscillations of the CCDD at fine timescales in Fig.~\ref{fig:CCDD_rect}. 
    Middle: decay time of the saturating curve of the CCDD at coarse timescales in Fig.~\ref{fig:CCDD_background}. 
    Bottom: decay time of the spin locking in Fig.~\ref{fig:spin_lock}.
    }
    \label{tab:summary_time_weak}
\end{table}

As discussed in Sec.~\ref{sec:spin_lock_c2_rect}, we attribute the subsequent decay in spin locking to the central line, while the initial decay arises predominantly from the sidebands. When a second drive is added to the spin locking for the transverse CCDD, the central line primarily induces oscillations, corresponding to the fine-timescale oscillations in Fig.~\ref{fig:CCDD_rect}(b). The envelope of these oscillations contributes to the initial decay observed on the coarse timescale in Fig.\ref{fig:CCDD_background}(a), while the remaining component gives rise to the background. In contrast, the sidebands are not effectively driven by the second drive and experience enhanced decay from the second drive. In summary, the oscillatory component of the CCDD stems mainly from the central line, while the non-oscillatory background originates from the undriven central line and the sidebands, whose decay dynamics are more sensitive to the initial states.

According to this argument, the spin locking with a weak drive is dominated by contributions from the central line. When a second drive is applied, this component gives rise to the oscillations in Fig.~\ref{fig:CCDD_rect}(d), which correspond to the initial decay observed on the corse timescale in Fig.~\ref{fig:CCDD_background}(b). In this case, the non-oscillatory background in Fig.~\ref{fig:CCDD_rect}(d) should originate from the central line that is not effectively driven by the second drive. However, this does not account for the residual saturation level $c_{\rm sl, w}$=-0.00196 in the inset of Fig.~\ref{fig:spin_lock}. We attribute this offset to spins that are not effectively prepared along the $x$-axis. For example, the states along the $z$-axis could evolve into a steady state under Rabi driving, which could account for the observed plateau in the inset. This steady state subsequently relaxes on the longitudinal $T_1$ timescale.  

\section{Fitting parameters}\label{sec:fit}
This section summarizes the fitting parameters for the data.

\subsection{Main}

\subsubsection{CCDD characterization}
The results of the Rabi sequence and the CCDD sequence in 
Fig.~\ref{fig_characterization}
are fit with a exponentially decaying cosine function: $A_{\Omega_n}\exp(-t/T_{\Omega_n})\cos(\Omega_nt)+ B_{\Omega_n}$. The resulting fitting parameters are shown in Table \ref{tab_CCDD}.

\begin{table}[h]
    \centering
    \caption{Fitting parameters for the results of Rabi sequence and CCDD sequence.}
    \label{tab_CCDD}
    \begin{tabular}{cccccc}
        \hline
        \hline
         & $\Omega_{n}$ & $T_{\Omega_n}$ &
        $A_{\Omega_n}$	&$p _{\Omega_n}$ & $C_{\Omega_n}$ \\
        \hline
        Rabi$(n=1)$	& 11.35 MHz & 197 ns 
        &	0.029 &	1.83 &-0.027 \\
        CCDD$(n=2)$	& 1.134 MHz	& 4.58 $\mu$s	
        & 0.025	 & -1.00 & -0.018 \\
        \hline
        \hline
    \end{tabular}
    \end{table}

\subsubsection{Magnetometry}
Transient dynamics under the Rabi magnetometry and the CCDD magnetometry in 
Fig.~\ref{fig_comparison}
are fit also with an exponentially decaying sinusoidal function: $A\exp(-t/T)\sin(\Omega t +p)+C$. The resulting parameters are shown in Table \ref{tab_comparison}.

\begin{table}[h]
    \centering
    \caption{Fitting paremeters for transient dynamics of direct Rabi magetometry and CCDD magnetometry.}
    \label{tab_comparison}
    \begin{tabular}{cccccc}
        \hline
        \hline
         & $\Omega/2\pi$ & $T$ & $A$	& $p$ & $C$ \\ 
        \hline 
        Rabi(weak)	 
        & 85.6 kHz	& 11.8 $\mu$s	& 0.0017 & 1.88 & -0.0021 \\
        Rabi(strong) 
        & 946 kHz	& 1.65 $\mu$s	& 0.0090 & 1.97 & -0.011 \\
        CCDD(weak)   
        & 20.5 kHz	& 140 $\mu$s	& 0.0026 & -1.19 &-0.00050 \\
        CCDD(strong) 
        & 168 kHz	& 32 $\mu$s	    & 0.011	 & 3.14 & -0.00010 \\
        \hline
        \hline
    \end{tabular}
\end{table}

\subsection{Appendix}

\subsubsection{Energy spectrum}
The energy spectrum in Appendix~\ref{sec:characteristics} was fitted with a sum of three Lorentzians:
\begin{equation}
    f(x) = \sum_{n=1,2,3}A_n \frac{\Delta\nu_n^2}{(x-f_n)^2+\Delta\nu_n^2} \nonumber.
\end{equation}
The resulting fitting parameters are summarized in Table \ref{tab:fit_params_energy_spectrum}.

\begin{table}[h]
    \caption{Fitting parameters for energy spectrum.}
    \label{tab:fit_params_energy_spectrum}

    \centering
    \begin{tabular}{cccc}
    \hline
    \hline
      $n$   &  $f_n$ [GHz]& $A_n$ & $\Delta\nu_n$[kHz] \\
    \hline

       +1   &2.705854     & $-2.71 \times 10^{-3}$& 385 \\        
       0    &2.708018     & $-3.35 \times 10^{-3}$& 415 \\        
       -1   &2.710167     & $-5.14 \times 10^{-3}$& 422 \\ 
    \hline
    \hline
       
    \end{tabular}
\end{table}

\subsubsection{Rabi}
The Rabi oscillations with strong and weak drives were fitted with an exponentially decaying cosine function:
\begin{equation}
    f(t)=A_{\Omega_1}\exp(-t/T_{\Omega_1})\cos(\Omega_1t). \nonumber
\end{equation}
The corresponding fitting parameters are summarized in Table \ref{tab:fit_Rabi}.
\begin{table}[h]
    \caption{Fitting parameters for Rabi oscillations.}
    \label{tab:fit_Rabi}

    \centering
    \begin{tabular}{cccc}
    \hline
    \hline
         &  $\Omega_1/(2\pi)$ & $A_{\Omega_1}$ & $T_{\Omega_1}$ \\
    \hline
       Strong   &11.35 MHz     & 0.0287 & 197 ns \\        
       Weak    &837 kHz    & 0.0071& 2.67 $\mu$s \\        
    \hline
    \hline   
    \end{tabular}
\end{table}

\subsubsection{CCDD}

\begin{figure*}[]
\includegraphics[width=\linewidth]{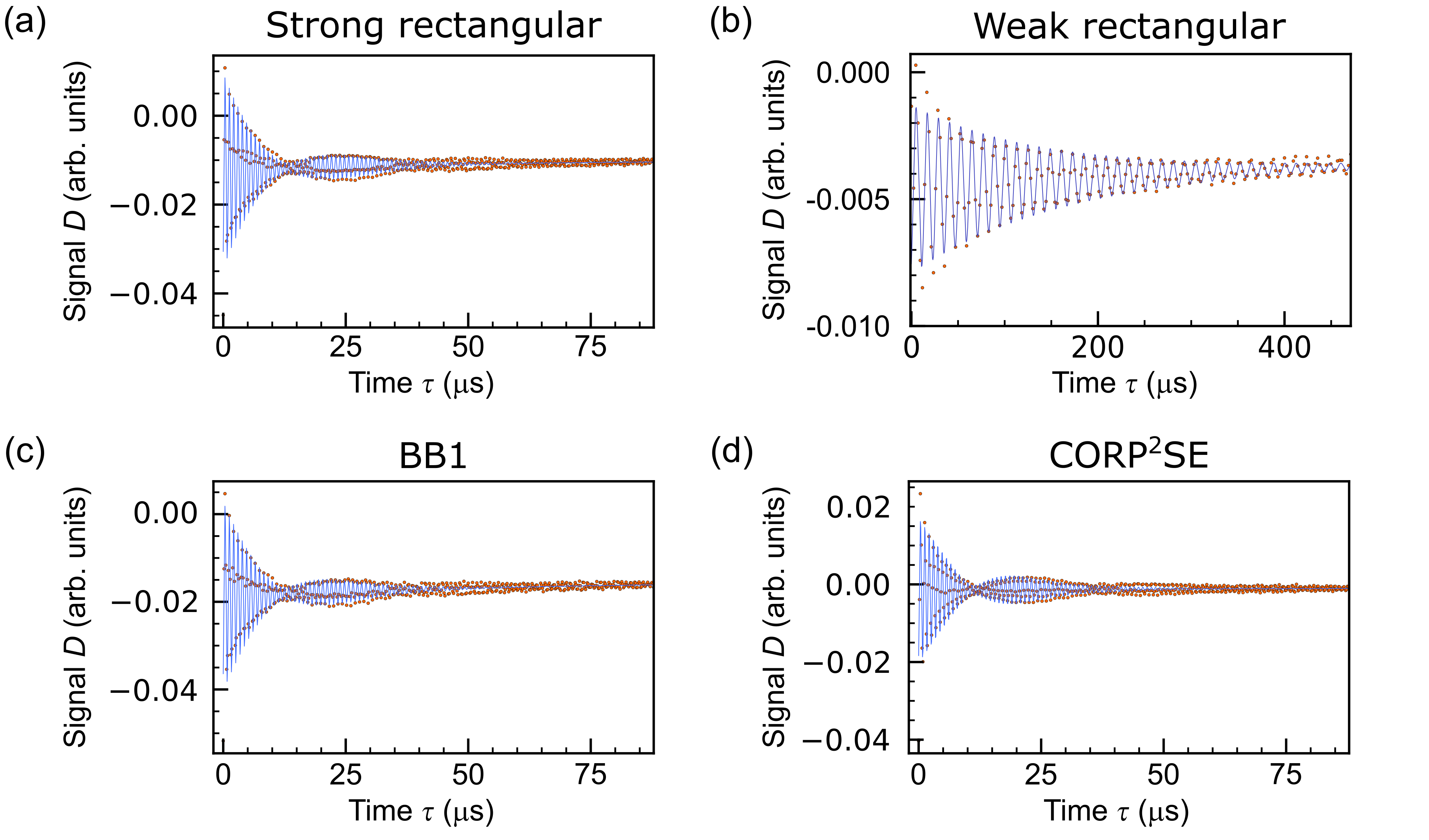}
\caption{\label{fig:CCDD_fit} Differential signals of the CCDD with corresponding fitting curves. Experimental data  are shown as orange circles, and the the fitted curves are overlaid in blue.}
\end{figure*}

As discussed in 
Appendix~\ref{sec:CCDD_rect} and \ref{sec:CCDD_CP},
the differential signal of the CCDD sequence is fitted with 
Eq.~\ref{eq:D_CCDD}.
The resulting fitting parameters are summarized in Table\ref{tab_fit_CCDD} and the corresponding fitting curves are shown in Fig.~\ref{fig:CCDD_fit}. 

This model captures the main features of the CCDD, but certain features are not handled. In particular, under strong drive with rectangular strong, BB1, and CORP$^2$SE pulses, long-lasting  oscillations are observed beyond $T_{\Omega_2}^{(1)}$ or $T_{\Omega_2}^{(2)}$. They should correspond to the line shape in the power spectrum density shown in the insets. Similar long-lasting oscillations are also present in Rabi oscillations. They may arise from combinations of different detuning and amplitude noise\cite{bertaina_experimental_2020, de_raedt_sustaining_2022}. Further study is required to understand their origin. On the other hand, the model does not capture the initial decay observed in the weak drive. Nonetheless, the current model provides a consistent framework for extracting the relevant timescales.

\FloatBarrier

\begin{table*}[]
    \centering
    \caption{Fitting parameters for CCDD.}
    \label{tab_fit_CCDD}
    \begin{tabular}{cccccccccccc}
    \hline
    \hline
     
    & $A_{\Omega_2}^{(1)}$ & $\Omega_2^{(1)}$ 
    & $T_{\Omega_2}^{(1)}$ & $\phi_{\Omega_2}^{(1)}$
    & $A_{\Omega_2}^{(2)}$ & $\Omega_2^{(2)}$ 
    & $T_{\Omega_2}^{(2)}$ & $\phi_{\Omega_2}^{(2)}$
    & $B_{\Omega_2}$ & $\tau_{\Omega_2}$ & $C_{\Omega_2}$ \\
    \hline
    Rect. strong
    & 0.0249   & 1.143 MHz	 & 12.0 $\mu$s	& -1.94	
    & 0.0140   & 1.130 MHz   & 15.5 $\mu$s  & 2.28
    & 0.0238   & 50.3 $\mu$s &	-0.0123 \\
    BB1	
    & 0.0195  & 1.144 MHz    & 12.1 $\mu$s	& -2.06
    & 0.00987 &	1.125 MHz	 & 16.0 $\mu$s	& 2.60
    & 0.00369 &	71.6 $\mu$s	 & -0.0187 \\
    CORP$^2$SE
    & 0.0125  &	1.146 MHz	 & 14.1 $\mu$s	& -2.65	
    & 0.00731 &	1.111 MHz	 & 15.6 $\mu$s	& 3.14
    & 0.000634& 39.5$\mu$s	         & -0.00163 \\
    Rect. weak
    & 0.00332 &	834.5 kHz    & 149 $\mu$s   & -2.65
    & -	      & -            &	-	        & -	
    & 0.0141  &	471 $\mu$s	 & -0.0461 \\
    \hline
    \hline
    \end{tabular}
\end{table*}

\subsubsection{Spin lock}
The signals from the spin-locking sequence in Appendix~\ref{sec:spin_lock} are fitted with Eq.~\ref{eq:D_spin_lock}
and the corresponding fitting parameters are shown in Table \ref{tab:fit_spin_lock}.

\begin{table}[h]
    \caption{Fitting parameters for spin locking.}
    \label{tab:fit_spin_lock}

    \centering
    \begin{tabular}{cccccc}
    \hline
    \hline
                &  $T_{\rm sl}^{(1)}$ & $B_{\rm sl}^{(1)}$ 
                & $T_{\rm sl}^{(2)}$ & $B_{\rm sl}^{(2)}$
                & $C_{\rm sl}$
    \\
    \hline
       Strong   
       & 36.4 $\mu$s & 0.0150 
       & 786 $\mu$s  & 0.0319 
       & $-7.21\times10^{-4}$\\        
       Weak(1.2 ms)   
       & 212 $\mu$s & 0.0039
       & -           &  -
       & -0.00196\\        
       Weak(17 ms)   
       & -           &  -
       & 3.87 $\mu$s & 0.00608
       & $3.70\times10^{-4}$
    \\        
       CORP$^2$SE   
       & 25.1 $\mu$s & 0.0206 
       & 888 $\mu$s  & 0.018 
       & $-9.16\times10^{-4}$
    \\        
    \hline
    \hline
    \end{tabular}
\end{table}
 
\FloatBarrier


%

\end{document}